# Liquidity Jump, Liquidity Diffusion, and Crypto Wash Trading

Qi Deng[1,2*]
and
Zhong-guo Zhou[3]

## Abstract

We develop a new framework to detect wash trading in crypto assets through real-time liquidity fluctuation.  We propose that short-term price jumps in crypto assets results from wash trading-induced liquidity fluctuation, and construct two complementary liquidity measures, liquidity jump (size of fluctuation) and liquidity diffusion (volatility of fluctuation), to capture the behavioral signature of wash trading.  Using US stocks as a benchmark, we demonstrate that joint elevation in both liquidity metrics indicates wash trading in crypto assets.  A simulated regulatory treatment that removes likely wash trades confirms this dynamic: it reduces liquidity diffusion significantly while leaving liquidity jump largely unaffected.  These findings align with a theoretical model in which manipulative traders amplify both the level and variance of price pressure, whereas passive investors affect only the level.  Our model offers practical tools for investors to assess market quality and for regulators to monitor manipulation risk on crypto exchanges without oversight.

Funding Source: The work was supported by Hubei University of Automotive Technology under Grant number BK202209; Hubei Provincial Bureau of Science and Technology under Grant number 2023EHA018.

1. College of Artificial Intelligence, Hubei University of Automotive Technology,
   167 West Checheng Road, Shiyan, Hubei Province, China
2. Cofintelligence Financial Technology Ltd., Hong Kong and Shanghai, China
3. Department of Finance, Financial Planning, and Insurance, Nazarian College of Business and Economics, California State University Northridge, CA, USA
*. Corresponding author: dq@huat.edu.cn; qi.deng@cofintelligence.com

# Liquidity Jump, Liquidity Diffusion, and Crypto Wash Trading

## 1. Introduction

Well-functioning financial markets rely on transparent price discovery and reliable liquidity. When manipulative traders engage in behaviors like wash trading, they distort prices and liquidity conditions, undermining market integrity and complicating the roles of investors and regulators, as well as the functioning of financial intermediation of liquidity provision. The trading of crypto assets remains largely unregulated and vulnerable to wash trading. Recent studies document widespread wash trading on unregulated crypto exchanges. Building on this evidence, we develop a framework to identify and measure wash trading through real-time liquidity diagnostics. Specifically, we introduce two complementary measures: liquidity jump and liquidity diffusion, which jointly capture the behavioral signature of wash trading. Le Pennec, Fiedler, and Ante (2021) analyze wash trading using web traffic and wallet data and estimate that over 90% of the reported trading volume on some crypto exchanges arises from such activities. Similarly, Aloosh and Li (2024) document Bitcoin wash trading across multiple exchanges, while Sifat, Tariq, and van Donselaar (2024) report comparable patterns in non-fungible token (NFT) transactions. Cong et al. (2023) classify crypto exchanges into regulated and unregulated groups and demonstrate that wash trading is common on unregulated platforms and nearly absent on regulated ones. Despite the methodological rigor of these studies, three key questions remain: 1) How can the magnitude and likelihood of crypto wash trading be adequately measured on a daily basis? 2) What indicators should investors monitor to make informed crypto trading decisions? and 3) What targeted actions should regulators take to combat wash trading.

Cong et al. (2023) find that the wash trading of a crypto asset is proportional to its price over the immediate horizon (during the same week), supporting the notion that manipulative traders use wash trading induce "short-term price jumps" and profit from them. We hypothesize that at least a portion of these short-term price jumps stem from liquidity variability caused by wash trading. In order to identify and then quantify the wash trading-induced liquidity fluctuation, we propose a model that decomposes asset liquidity into two components: "liquidity jump" and "liquidity diffusion." Liquidity jump, defined as the ratio of regular return to liquidity-adjusted return, measures the size of liquidity fluctuation, and liquidity diffusion, defined as the ratio of regular volatility to liquidity-adjusted volatility, reflects the intraday volatility of liquidity fluctuation.

We contend that liquidity jump and liquidity diffusion serve as proxies for the scale and probability of wash trading, respectively. To contextualize this intuition, we present a simple theoretical model in which manipulative traders amplify both liquidity measures, whereas passive investors influence only the liquidity jump. The model provides economic grounding for our empirical detection framework and supports the interpretation of our jump-diffusion measures as behavioral signals of manipulation. By comparing that trading data from the regulated US stock market (where wash trading is non-existent) with that from the unregulated crypto markets (where wash trading is common), we observe that manipulative traders tend to generate elevated liquidity diffusion to induce pronounced liquidity jump, thereby profiting from the resultant price dislocations. The joint occurrence of high liquidity diffusion and high liquidity jump is posited as an indicator of wash trading, whereas large-volume trades that exhibit high liquidity jump but low liquidity diffusion are more likely legitimate high-demand transactions. A simulated regulatory treatment supports the interpretation.

From the perspective of investors, both liquidity jump and liquidity diffusion provide specific information to help them make informed investments in crypto assets (perceivably) infested with wash trading. Our analysis of the top 10 market-cap crypto assets indicates that wash trading is more concentrated among smaller-cap assets and relatively uncommon among larger assets such as Bitcoin (BTC) that trade in established though unregulated exchanges (e.g., Binance). This observation suggests that investing in high-market-cap crypto assets may provide portfolio diversification. From the perspective of regulators, the liquidity jump and liquidity diffusion framework provides a systematic and reliable approach for detecting liquidity fluctuation caused by wash trading, which enables them to take appropriate actions to combat wash trading. We contribute to the literature on liquidity measurement, market manipulation, and surveillance methodology. Although we only use crypto assets to exemplify the usefulness and utilities of our model, our model is generalizable to any asset class exhibiting high-frequency liquidity variability.

The rest of the paper proceeds as follows. Section 2 reviews literature on liquidity modeling. Section 3 introduces the asset-level liquidity-adjusted return and volatility model that produces liquidity jump and liquidity diffusion. Section 4 provides descriptive statistics and visualizations of the datasets (with the US stocks, crypto assets without and with a "simulated" treatment on wash trading) with an analysis of their distributions and implications, and interprets relevancy of liquidity jump and liquidity diffusion to wash trading. Section 5 concludes the paper with a discussion of the findings.

## 2. Literature Review

We propose that wash trading initiators attempt manipulative trades to stimulate short-term price jumps in order to profit, and that these short-term price jumps are manifestations of liquidity fluctuations driven by such manipulative activities. In this section, we review the relevant literature on liquidity proxies, liquidity fluctuation, liquidity volatility, and models of assets with extreme liquidity variability to identify research gaps in the context of revealing the connections between liquidity fluctuation (both size and probability) and crypto wash trading.

**2.1 Liquidity Proxies**

Fong, Holden, and Trzcinka (2017) identify a multitude of high-quality liquidity proxies and find that the daily version of the Closing Percent Quoted Spread (Chung and Zhang, 2014) is the best daily percent-cost proxy, while the daily version of Amihud (2002) is the best cost-per-dollar-volume proxy. Amihud and Mendelson (1986) define an asset's illiquidity as the cost of immediate execution (trading friction) and use the bid-ask spread as a general measure of illiquidity, though some large trades may occur outside the bid-ask spread (Chan and Lakonishok, 1995; Keim and Madhavan, 1996). While the bid-ask spread explains the returns of NASDAQ, it does not fully capture those of NYSE and AMEX (Eleswarapu, 1997). Consequently, some studies suggest using price impact (return) as an alternative liquidity measure (Brennan and Subrahmanyam, 1996; Amihud, 2002; Pástor and Stambaugh, 2003; Acharya and Pedersen, 2005). Given that we use minute-interval trading data without order book (bid-ask) information, the daily version of Amihud (2002) is a natural starting point for constructing our own liquidity measures. Specifically, we propose a model that derives liquidity measures (jump and diffusion) from asset returns (regular and liquidity-adjusted), based on a minute-level liquidity ratio similar to that of Amihud (2002).

**2.2 Liquidity Jump**

The liquidity jump (a proxy for the size of liquidity fluctuation) captures the potential gains of manipulative traders. It builds on economic theories in market microstructure, behavioral finance, and liquidity premium, particularly the well-established idea that low-liquidity assets tend to exhibit higher liquidity fluctuations. Market microstructure and behavioral finance theories explain how market design influences price formation, trading behavior, and liquidity. Kyle (1985) shows how informed traders exploit uninformed liquidity traders, leading to wider bid-ask spreads and greater liquidity fluctuations. In low-liquidity markets, the risk of adverse selection is higher, prompting market makers to widen spreads to protect against informed trading, resulting in greater liquidity variability. Glosten and Milgrom (1985) demonstrate how market makers adjust spreads in response to the probability of trading with informed traders. Similarly, Ho and Stoll (1981) show how market makers manage inventory risk by adjusting quotes in response to inventory imbalances, which are more pronounced in illiquid markets, causing greater liquidity fluctuation. Easley and O'Hara (1987) highlight the impact of trade size on liquidity, with large trades in illiquid markets often causing significant price movements and liquidity fluctuations.

The liquidity premium theory explains the pricing of assets in relation to liquidity risk. Amihud and Mendelson (1986) propose that exogenous transaction costs and trading horizons are fundamental sources of illiquidity and argue that the additional return required to compensate for increased spread is higher for low-spread stocks. Amihud, Mendelson, and Pedersen (2005) further show that time-varying illiquidity impacts stock prices, requiring investors to demand compensation for liquidity risk. They present and fine-tune a number of pricing models to reflect such a liquidity premium, which is typically modeled as an additional term in the expected return equation. We extend their models by incorporating liquidity premium directly into liquidity-adjusted return, where the ratio of regular return to liquidity-adjusted return is the liquidity jump.

## 2.3 Liquidity Diffusion

Time-varying liquidity affects volatility as it affects prices (Amihud, Mendelson, and Pedersen, 2005). Petkova, Akbas, and Armstrong (2011) find that stocks with higher liquidity variability command higher expected returns, suggesting that investors dislike liquidity volatility due to potential liquidity downturns. Conversely, Chordia, Subrahmanyam, and Anshuman (2001) show a strong negative relationship between liquidity volatility and expected returns, with Pereira and Zhang (2010) finding that stocks with higher liquidity volatility tend to earn lower returns. Amiram et al. (2019) reconcile these views by decomposing total volatility into a jump component and a diffusion component, observing a positive relationship between jump volatility and illiquidity and a negative relationship between diffusive volatility and illiquidity, suggesting different effects on liquidity risk premium for the two volatility components. Inspired by their work, we propose decomposing liquidity fluctuation into two components: liquidity jump and liquidity diffusion, which in turn help identify wash trades and distinguish wash trades from legitimate high-liquidity trades at the same time.

Andersen, Bollerslev, and Meddahie (2011) extend the diffusive volatility model to include market microstructure noise (transient distortions in observed prices due to bid-ask bouncing, order imbalances, and latency), in which liquidity is factored implicitly. Bollerslev and Todorov (2023) show that market price jumps command a different risk premium than continuous price moves and develop short-time, risk-neutral variance expansions to delineate price and variance risks. They find that simultaneous jumps in the price and the stochastic volatility command a sizeable risk premium. Their approach supports our framework of decomposing liquidity into jump and diffusion components, suggesting that manipulative traders profit from increases in both.

The liquidity diffusion (a proxy for liquidity volatility) reveals manipulative trading patterns and inherently grounded on the economic theory of market microstructure theory, particularly microstructure noise. Aït-Sahalia, Mykland, and Zhang (2005) demonstrate the importance of disentangling true price movements from microstructure noise to avoid biased high-frequency volatility estimates. Transient liquidity crunches, often driven by manipulative trades, reflect noise rather than fundamental value shifts, complicating volatility forecasts. These findings underscore the need for microstructure noise models integrated with liquidity information in capturing liquidity-induced volatility (e.g., Menkveld, 2013; Tripathi and Dixit, 2020) and in particular, call for modeling noises with integrated liquidity information that the liquidity diffusion provides.

**2.4 Models of Assets with Extreme Liquidity Variability**

Most existing liquidity volatility models are rarely tested on assets with extreme liquidity variability because regulated markets generally exhibit stable liquidity patterns. However, the rapid rise of unregulated crypto markets offers a unique opportunity to study high-liquidity variability assets. Cong et al. (2023) categorize 29 centralized crypto exchanges into regulated and unregulated groups, finding that wash trades account for more than 70% of reported volumes on unregulated exchanges, necessitating purposely designed models for crypto assets. Using millisecond-level data, Manahov (2021) shows that extreme price movements in major crypto assets demand liquidity premium and reflect herding behavior in up markets. Shen, Urquhart, and Wang (2022) identify intraday trading momentum in Bitcoin, driven by liquidity provision. These studies suggest that a likely trading strategy for manipulative traders is to first create high liquidity volatility (diffusion) through intentionally placed bid-ask trades and then exploit herding behavior and momentum (jump) to profit from unsuspected investors. Our liquidity measures are particularly suited for modeling the liquidity risks associated with wash-traded crypto assets.

**2.5 Liquidity Decomposition and Research Gaps**

In addition to some research mentioned above (e.g., Amiram et al., 2019; Bollerslev and Todorov, 2023), other earlier work, such as Amihud (2002) and Hou and Moskowitz (2005) shows that liquidity can vary across different market conditions and asset classes, which, in essence, discuss the cross-sectional liquidity variability (jump). Gabaix et al. (2006) find the evidence of clustering in jumps and that extreme liquidity is not purely random but is driven by common factors or dynamics in the market, which contemplates the temporal liquidity variability (diffusion). Another stream of research, such as by Datar, Naik, and Radcliffe (1998), Chordia and Subrahmanyam (2004), and Bekaert, Harvey and Lundblad (2007), shows that the level of liquidity variability is generally associated with transaction costs, bid-ask spreads, and market depth, touching upon liquidity diffusion in the context of market microstructure. Roll (1984), Huberman and Halka (2001), and Aït-Sahalia, Mykland and Zhang (2005) share a broader view on a crucial role that liquidity plays towards market efficiency, pricing dynamics, and risk management in continuous-time processes, revealing both cross-sectional and temporal nature of liquidity jump and liquidity diffusion, respectively.

However, there are still two major gaps in the abovementioned literature: 1) there is no explicit attempt to actually divide liquidity into a jump component and a diffusion component, and therefore the two-way microstructure-level interaction between liquidity and trading cannot be further fine-tuned; and 2) most studies have an implicit assumption that the assets under investigation are strictly-regulated and therefore don't exhibit extreme liquidity variability and are not subject to wash trading. Both gaps render their models inadequate in dealing with crypto wash trading. To our best knowledge, this paper is the first attempt to explicitly decompose asset liquidity into liquidity jump and liquidity diffusion, and to connect these liquidity measures to the size and probability of crypto wash trading.

## 3. Liquidity Jump and Liquidity Diffusion

## 3.1 Liquidity Jump and Liquidity Diffusion – minute-level

Imagine that a risk-averse investor had a choice between two assets, A and B, with identical expected return and volatility, yet A had a much higher trading volume/amount than B, the investor most likely would choose A, particularly if the investor's horizon were not indefinite. The intuitive reason for this choice is that the perceived volatility of A is lower than that of B. This choice is consistent with Amihud, Mendelson and Pedersen (2005). Using the minute-level trading data (crypto trading data from the largest crypto asset exchange Binance API and stock trading data from the Polygon.io API), we model this perceived and unobservable volatility as a minute-level liquidity-adjusted variance $\sigma^2{}_T^\ell$ at equilibrium for time-period $T$ (a 24-hour/1440-minute trading day for crypto assets and a 6.5-hour/390-minute trading day for US stocks), with the variance of each minute being adjusted by an illiquidity factor similar to that of Amihud (2002). The expression of $\sigma^2{}_T^\ell$ is given as follows ($\tau$ is minute-level time index for minute $\tau$):

$$\sigma^2{}_T^\ell = \frac{1}{T}\sum_{\tau=1}^{T} \eta_T \frac{|r_\tau|/\overline{|r_\tau|}}{A_\tau/\overline{A_\tau}} (r_\tau - \overline{r_\tau})^2 = \frac{1}{T}\sum_{\tau=1}^{T} \eta_T \ell_\tau (r_\tau - \overline{r_\tau})^2 = \frac{1}{T}\sum_{\tau=1}^{T} \ell_T (r_\tau - \overline{r_\tau})^2 \qquad (1)$$

$$\Rightarrow \sigma^2{}_T^\ell = \frac{1}{T}\sum_{\tau=1}^{T}\left(\sqrt{\eta_T \frac{|r_\tau|/\overline{|r_\tau|}}{A_\tau/\overline{A_\tau}}}\, r_\tau - \sqrt{\eta_T \frac{|r_\tau|/\overline{|r_\tau|}}{A_\tau/\overline{A_\tau}}}\, \overline{r_\tau}\right)^2 = \frac{1}{T}\sum_{\tau=1}^{T}\left(\sqrt{\eta_T \ell_\tau}\, r_\tau - \sqrt{\eta_T \ell_\tau}\, \overline{r_\tau}\right)^2 = \frac{1}{T}\sum_{\tau=1}^{T}\left(\sqrt{\ell_T}\, r_\tau - \sqrt{\ell_T}\, \overline{r_\tau}\right)^2$$

where:

$$\ell_\tau = \frac{|r_\tau|/\overline{|r_\tau|}}{A_\tau/\overline{A_\tau}}$$

$$\ell_T = \eta_T \frac{|r_\tau|/\overline{|r_\tau|}}{A_\tau/\overline{A_\tau}} = \eta_T \ell_\tau$$

subject to:

$$\sum_{\tau=1}^{T} \eta_T \frac{|r_\tau|/\overline{|r_\tau|}}{A_\tau/\overline{A_\tau}} = \sum_{\tau=1}^{T} \eta_T \ell_\tau = \sum_{\tau=1}^{T} \ell_T = T$$

$$\Rightarrow \eta_T = \frac{T}{\sum_{\tau=1}^{T} \frac{|r_\tau|/\overline{|r_\tau|}}{A_\tau/\overline{A_\tau}}} = \frac{T}{\sum_{\tau=1}^{T} \ell_\tau}$$

*where:*
*$r_\tau$ is the observed return at minute $\tau$, $|r_\tau|$ is its absolute value, $\overline{|r_\tau|}$ is its arithmetic average in that day,*
*$A_\tau$ is the dollar amount traded at minute $\tau$, $\overline{A_\tau}$ is its arithmetic average in day $T$,*
*$\eta_T$ is the daily normalization factor on day $T$ and is a constant for day $T$,*
*$T$ = 1440 minutes (24 hours) or 390 minutes (6.5 hours) in a crypto or US stock trading day, respectively*

The term $\ell_\tau$ is essentially a normalized (~1) minute-level version of Amihud (2002), with both numerator and denominator of $\ell_\tau$ being normalized with regard to its respective parameters, the absolute return and trading amount ($|r_\tau|$ and $A_\tau$), and $\ell_T$ is a further normalized $\ell_\tau$ with a daily normalization factor $\eta_T$. The term $\ell_\tau$ reflects the variability of a particular asset's intraday liquidity, which is useful for intraday trading; the term $\ell_T$ provides direct daily liquidity comparison between assets, which is useful for portfolio optimization.

Furthermore, in Equation 1, $\sigma^2{}_T^\ell$ is the variance of an unobservable minute-level liquidity-adjusted return at equilibrium, $r_\tau^\ell$:

$$\sigma^2{}_T^\ell = \frac{1}{T}\sum_{\tau=1}^{T}(r_\tau^\ell - \overline{r_\tau^\ell})^2 \qquad (2)$$

where: $r_\tau^\ell$ is the liquidity-adjusted return at minute $\tau$, $\overline{r_\tau^\ell}$ is its arithmetic average in day T.

By equating the right-hand side of Equations 1 and 2, we connect an asset's observed return $r_\tau$ and its unobservable liquidity-adjusted counterpart $r_\tau^\ell$ (1st order approximation) as follows[1]:

$$r_\tau^\ell = \sqrt{\eta_T \frac{|r_\tau|/\overline{|r_\tau|}}{A_\tau/\overline{A_\tau}}} r_\tau = \sqrt{\eta_T \ell_\tau} r_\tau = \sqrt{\ell_T} r_\tau = \frac{1}{\beta_{r_\tau}^\ell} r_\tau \qquad (3)$$

$$\Rightarrow r_\tau = \beta_{r_\tau}^\ell r_\tau^\ell$$

$$\Rightarrow \beta_{r_\tau}^\ell = \frac{r_\tau}{r_\tau^\ell} = \frac{1}{\sqrt{\eta_T \frac{|r_\tau|/\overline{|r_\tau|}}{A_\tau/\overline{A_\tau}}}} = \frac{1}{\sqrt{\eta_T \ell_\tau}} = \frac{1}{\sqrt{\ell_T}}$$

where:
$r_\tau$ is observed return at minute $\tau$, $r_\tau^\ell$ is the liquidity-adjusted return at minute $\tau$, $\beta_{r_\tau}^\ell$ is the liquidity factor at minute $\tau$.

We structure $\beta_{r_\tau}^\ell$ in such a way that it is a unitless liquidity measure and normalized (~1), and that when liquidity is high (amount is high), $r_\tau$ is greater than $r_\tau^\ell$, while liquidity is low (amount is low), $r_\tau$ is smaller than $r_\tau^\ell$. This reflects the intuition of a liquidity premium, that a perfectly liquid (frictionless, $A_\tau \to \infty$) stock generates no liquidity-adjusted return ($r_\tau^\ell \to 0$), and that a perfectly illiquid (infinite friction, $A_\tau \to 0$) stock earns an infinite liquidity-adjusted return ($r_\tau^\ell \to \infty$).[2] We thus refer $\beta_\tau^\ell$ as the minute-level "liquidity fluctuation Beta" given as:

$$\beta_{r_\tau}^\ell = \frac{1}{\sqrt{\eta_T \ell_\tau}} = \frac{1}{\sqrt{\ell_T}} \subset \begin{cases} > 1; high\ minute\ liquidity\ fluctuation \\ = 1; equilibrium\ minute\ liquidity\ fluctuation \\ < 1; low\ minute\ liquidity\ flutuation \end{cases} \quad (4)$$

---

[1] Note that although we define $r_\tau^\ell = \sqrt{\eta_T \ell_\tau} r_\tau$, in general $\overline{r_\tau^\ell} = \sqrt{\eta_T \ell_\tau} \overline{r_\tau}$ does not hold. However, its 1st order approximation is acceptable in literature.

Furthermore, $\sigma^2{}_T$ is the variance of the observed regular minute-level return $r_\tau$, thus we also introduce a minute-level "liquidity volatility Beta," $\beta_{\sigma_\tau}^\ell$, which reflects the volatility of liquidity for time-period T, defined as follows:

$$\beta_{\sigma_\tau}^\ell = \sigma_T/\sigma_T^\ell \subset \begin{cases} > 1; high\ minute\ liquidity\ volatility \\ = 1; equilibrium\ minute\ liquidity\ volatility \\ < 1; low\ minute\ liquidity\ volatility. \end{cases} \quad (5)$$

Together, $\beta_{r_\tau}^\ell$ and $\beta_{\sigma_\tau}^\ell$ form the "minute-level liquidity Beta" pair, which enables us to further derive the daily-level liquidity fluctuation and liquidity volatility measures.

### 3.2 Liquidity Jump and Liquidity Diffusion – daily-level

We then construct the daily-level returns by aggregating the intraday (minute-level) returns. Both the regular daily return and the realized but unobservable daily liquidity-adjusted returns on day $t$ are given as ($t$ is daily-level time index for day $t$):

$r_t = (1 + r_\tau)^T - 1$

$r_t^\ell = (1 + r_\tau^\ell)^T - 1$

The "daily liquidity fluctuation Beta" or liquidity jump $\beta_{r_t}^\ell$ for day $t$ is thus defined as:[3]

$$\beta_{r_t}^\ell = |r_t/r_t^\ell| \subset \begin{cases} > 1; high\ daily\ liquidity\ fluctuation \\ = 1; equilibrium\ daily\ liquidity\ fluctuation \\ < 1; low\ daily\ liquidity\ fluctuation \end{cases} \quad (6)$$

The realized and unobservable daily intraday (minute-level) variance for time-period $T$ is:

---

[2] Note that $\ell_\tau$ is zero if $r_\tau$ is zero in that minute, which makes $\beta_{r_\tau}^\ell$ infinite. To correct this singularity, we assign a random number two magnitudes ($10^{-2}$) smaller than $\bar{r}_\tau$ to each $r_\tau$ with a zero value.

[3] Unlike $\beta_{r_\tau}^\ell$, it is possible that $r_t/r_t^\ell$ is negative, therefore $\beta_{r_t}^\ell$, a positive value, is defined as $|r_t/r_t^\ell|$.

$$\sigma^{2\ell}_t = T\sigma^{2\ell}_T$$

In addition to $\beta^\ell_{r_t}$, we also introduce a "intraday liquidity volatility Beta" or liquidity diffusion $\beta^\ell_{\sigma_t}$, which reflects the liquidity volatility on day *t*, defined as:

$$\beta^\ell_{\sigma_t} = \sigma_t/\sigma^\ell_t \subset \begin{cases} > 1; \text{high daily liquidity volatility} \\ = 1; \text{equilibrium daily liquidity volatility} \\ < 1; \text{low daily liquidity volatility} \end{cases} \qquad (7)$$

The proposed liquidity jump $\beta^\ell_{r_t}$ and liquidity diffusion $\beta^\ell_{\sigma_t}$ are rooted in established financial theories. Specifically, these measures draw on insights from market microstructure theory, behavioral finance, and liquidity risk-premium theory. They also reflect the inherent characteristics of assets with different liquidity levels, capturing how liquidity behaves for assets ranging from highly illiquid to very liquid. The liquidity jump $\beta^\ell_{r_t}$ captures sudden and discontinuous changes in the daily liquidity level of an asset. Assets with low liquidity (e.g. certain crypto assets) tend to have more days with high liquidity jump ($\beta^\ell_{r_t} > 1$), indicating frequent large swings or jumps in daily liquidity. In contrast, highly liquid assets (e.g. large-cap U.S. stocks) more often exhibit low liquidity jump ($\beta^\ell_{r_t} < 1$), reflecting relatively stable day-to-day liquidity with fewer abrupt changes. These patterns align with market microstructure and behavioral finance theories, which suggest that less-liquid assets experience greater volatility in liquidity.

The liquidity diffusion $\beta^\ell_{\sigma_t}$ reflects the intraday volatility of liquidity conditions. An asset with stable trading volume (e.g., U.S. stock) typically has low liquidity diffusion ($\beta^\ell_{\sigma_t} < 1$), indicating low intraday variability in liquidity. Conversely, an asset with erratic or sporadic trading volume (e.g., crypto assets) induced by manipulative trades sees a higher value ($\beta^\ell_{\sigma_t} > 1$), signaling high intraday liquidity variability. This behavior is consistent with liquidity risk-premium theory, which emphasizes that greater variability in liquidity (higher liquidity risk) commands a higher expected return in asset pricing. In addition, $\beta^{\ell}_{\sigma_t}$ captures the frequency and intensity of liquidity shocks, a key concept in market microstructure. High values of $\beta^{\ell}_{\sigma_t}$ ($> 1$) indicate frequent and severe liquidity shocks that can lead to large price movements, while low values $\beta^{\ell}_{\sigma_t}$ ($< 1$) suggest infrequent and mild liquidity shocks. This aligns with market microstructure theory that liquidity shocks tend to increase price volatility and thus require a higher return as compensation which is especially relevant in the context of crypto wash trading.

Together, $\beta^{\ell}_{r_t}$ and $\beta^{\ell}_{\sigma_t}$ form a "liquidity Beta" pair that provides a more comprehensive view of liquidity risk on daily basis. In particular, examining the combination of $\beta^{\ell}_{r_t}$ and $\beta^{\ell}_{\sigma_t}$ provides useful tools to researchers to study the interplay between sudden liquidity change (jump) and gradual changes in liquidity (diffusion), helping them understand the behavioral patterns of manipulative traders. From the perspective of regulators, the interaction between $\beta^{\ell}_{r_t}$ and $\beta^{\ell}_{\sigma_t}$ can be monitored to track liquidity dynamics and detect manipulative trading practices in real time. From a liquidity provision standpoint, market makers can examine the prevailing liquidity conditions inferred from the liquidity Beta pair to adjust their quotes to improve market efficiency. For the benefit of investors, $\beta^{\ell}_{r_t}$ and $\beta^{\ell}_{\sigma_t}$ can be incorporated into their investment strategies to minimize transaction costs and optimize execution timing. Finally, that both liquidity measures leverage high-frequency data allows market participants to capture granular intraday liquidity dynamics. By incorporating information from within a trading day, these measures provide a detailed picture of liquidity behavior that traditional daily or lower-frequency metrics might miss. This high-frequency insight can be crucial for fine-tuning trading strategies, improving regulatory surveillance, and enhancing our overall understanding of liquidity risk in financial markets.

## 4. Dataset, Descriptive Statistics, and Discussions

## 4.1 Dataset

We use the US stock trading data to establish a benchmark to assess the utilities of liquidity jump and liquidity diffusion as tools for identifying wash trades. We collect minute-level trading data of all 1,503 constituent stocks of the SP500 (large cap), SP400 (mid cap) and SP600 (small cap) indices from the Polygon.io API. From each index, we pick the five largest stocks in terms of market cap, thus we select 15 stocks from the three indices.[4] All 15 stocks have at least ten years of complete historical data (July 28, 2014 to February 10, 2025 with 2,652 trading days). The selected stocks, in alphabetical order with their ticker symbols are AAPL, AMZN, ATI, CMA, CRS, EME, GOOG, IBKR, LII, MLI, MSFT, NVDA, TPL, VFC and WSO. We construct and calculate minute-level return and variance, both regular and liquidity-adjusted. We further aggregate the minute-level data into daily return and (intraday) volatility, both regular ($r_t$ and $\sigma_t$) and liquidity-adjusted ($r_t^\ell$ and $\sigma_t^\ell$), as well as the daily liquidity jump $\beta_{r_t}^\ell$ and daily liquidity diffusion $\beta_{\sigma_t}^\ell$.[5] We report the descriptive statistics of $\beta_{r_t}^\ell$ and $\beta_{\sigma_t}^\ell$ for all 15 US stocks in Table 1.

We then collect tick-level trading data of the top 15 non-stable-coin crypto assets (prices measured with their trading pairs with Tether or USDT)[6] in their market caps from the Binance API[7] with at least four years of complete historical data (October 15, 2020 to February 7, 2025 with 1,577 trading days). We discard crypto assets that are either platform gas/incentive tokens (e.g., BNB) or task facility/incentive tokens (e.g., LINK), of which liquidity provision is affected by non-trading activities and therefore distorted, ending up with 10 crypto assets. The selected crypto assets, in their alphabetical order with ticker symbols are ADA, AVAX, BCH, BTC, ETC, ETH, LTC, SOL, UNI and XRP. We calculate the daily $\beta^{\ell}_{r_t}$ and $\beta^{\ell}_{\sigma_t}$ for crypto assets with the same steps as those for stocks, and report their descriptive statistics for crypto assets in Table 2.

---

[4] We obtain the market caps of all 1,503 stocks from the Yahoo Finance API on October 12, 2024 at the market close.
[5] We cap the value of both $\beta_{r_t}^\ell$ and $\beta_{\sigma_t}^\ell$ to 10 in order to avoid extreme outliers that may bias our analysis.
[6] USDT is a "stable coin" pegged to the US Dollar, and for our purpose of portfolio construction it is regarded as a "risk-free" asset with a 0% interest rate.
[7] https://coinmarketcap.com, accessed on October 12, 2024.

## 4.2 Descriptive Statistics and Distribution of Stock $\beta^{\ell}_{r_t}$ and $\beta^{\ell}_{\sigma_t}$

Panel A of Table 1 summarizes the descriptive statistics of liquidity jump $\beta^{\ell}_{r_t}$ for all stocks. For $\beta^{\ell}_{r_t}$, the mean is between 0.44 (TPL) and 1.39 (GOOG) and the median ranges from 0.20 (TPL) to 0.87 (AMZN). The number of days with the maximum value of $\beta^{\ell}_{r_t}$ (=10) is in the range of 20 or 0.75% (TPL) to 65 or 2.45% (GOOG) of the total number of 2,652 trading days, which suggests that the trades with extreme liquidity are scarce among stocks.[8] The number of days with $\beta^{\ell}_{r_t} \geq 1$ spans from 187 or 7.05% (TPL) to 1,015 or 38.27% (AMZN). We observe that, the five largest market-cap stocks, AAPL, AMZN, GOOG, MSFT and NVDA, also have the highest $\beta^{\ell}_{r_t}$ in both mean and median (mean: 1.20-1.39, median: 0.82-0.87), and the highest number of days with high $\beta^{\ell}_{r_t}$ (804 or 30.32% to 1,015 or 38.27%). Panel B of Table 1 provides descriptive statistics of liquidity diffusion $\beta^{\ell}_{\sigma_t}$ for all stocks. For $\beta^{\ell}_{\sigma_t}$, the mean is between 0.22 (TPL) and 0.79 (AMZN) and the median ranges from 0.21 (TPL) to 0.80 (AMZN). The number of days with $\beta^{\ell}_{\sigma_t} \geq 1$ is between 0 (CRS, EME, MLI, TPL and WSO) and 29 or 1.09% (NVDA), suggesting that liquidity volatility is low for stocks. Again, the largest market-cap stocks (AAPL, AMZN, GOOG, MSFT and NVDA) also have the highest $\beta^{\ell}_{\sigma_t}$ in both mean and median (mean: 0.76-0.79, median: 0.76-0.80), and the highest number of days with $\beta^{\ell}_{\sigma_t} \geq 1$ (6 or 0.23% to 29 or 1.09%).

---

[8] For the purpose of being concise, from this point and on we just use a number and a percentage (e.g., 20 or 0.75%) to represent the longer phase (e.g., 20 days or 0.75% out of the total number of 2,652 trading days).

Combining the descriptive statistics of $\beta^\ell_{r_t}$ (Panel A) and $\beta^\ell_{\sigma_t}$ (Panel B), we establish that the US stocks display a pattern of very low level of liquidity diffusion (the maximum number of days with $\beta^\ell_{\sigma_t} \geq 1$ is 1.09% for NVDA) and reasonable low level of liquidity jump (the number of days with $\beta^\ell_{r_t} \geq 1$ between 7.05% and 38.27%) on daily level. As the US stocks are heavily regulated and therefore virtually immune to wash trading, with this rather strong and indicative empirical evidence, we conjecture that wash trading can only occur on days when both liquidity measures are high ($\beta^\ell_{\sigma_t} \geq 1$ and $\beta^\ell_{r_t} \geq 1$).

From Panel C of Table, we find that there are between 0 and 15 or 0.57% (NVDA) trading days with both high liquidity diffusion ($\beta^\ell_{\sigma_t} \geq 1$) and high liquidity jump ($\beta^\ell_{r_t} \geq 1$). It is worth mentioning that the top five stocks in the number of trading days with high liquidity measures, AAPL (13 or 0.49%), AMZN (11 or 0.41%), GOOG (4 or 0.15%), MSFT (10 or 0.38%) and NVDA (15 or 0.57%), are also the top five stocks in market cap. This suggests that the demand for these stocks is high, and the pressure for legitimate high-volume trades creates both higher liquidity fluctuation and liquidity volatility compared to other stocks. We will utilize this fact to gauge the level of wash trading among crypto assets in the next several subsections.

### 4.3 Descriptive Statistics and Distribution of Crypto $\beta^\ell_{r_t}$ and $\beta^\ell_{\sigma_t}$

Panel A of Table 2 summarizes the descriptive statistics of liquidity jump $\beta^\ell_{r_t}$ for all the selected crypto assets without treatment on wash trading. The mean is between 1.54 (BTC) and 1.99 (BCH, UNI) and the median is between 0.85 (XRP) and 1.09 (SOL). The number of days with high liquidity jump ($\beta^\ell_{r_t} \geq 1$) ranges from 652 or 41.34% (XRP) to 868 or 55.04% (SOL). It is worth mentioning that the three largest crypto assets in both market cap and liquidity, BTC, ETH and XRP, have the lowest $\beta^\ell_{r_t}$ in both mean and median (mean: 1.54, 1.63, 1.57, median:

0.92, 1.01, 0.85, respectively), and the lowest number of days with high $\beta_{r_t}^{\ell}$ (708 or 44.90%, 800 or 50.73%, 652 or 41.34%, respectively).[9] Panels B of Table 2 summarize the descriptive statistics of liquidity diffusion $\beta_{\sigma_t}^{\ell}$. The mean is between 0.82 (BTC) and 1.15 (BCH, UNI) and the median between 0.82 (BTC) and 1.06 (BCH). The number of days with high $\beta_{\sigma_t}^{\ell}$ ($\geq 1$) ranges from a low of 15 or 0.95% (BTC) to a high of 968 or 61.38% (BCH). Again, the three largest crypto assets (BTC, ETH and XRP) have the lowest number of days with high $\beta_{\sigma_t}^{\ell}$ ($\geq 1$) (15 or 0.95%, 77 or 4.88%, 124 or 7.86%, respectively).

Panel C of Table 2 captures the number of days with high $\beta_{\sigma_t}^{\ell}$ ($\geq 1$) and high $\beta_{r_t}^{\ell}$ ($\geq 1$), with a rather wide span of 8 or 0.51% (BTC) to 545 or 34.56% (BCH). Based upon our conjuncture (Subsection 4.2) that wash trading exhibits both high $\beta_{r_t}^{\ell}$ and high $\beta_{\sigma_t}^{\ell}$, we observe that the selected crypto assets demonstrate rather different levels of wash trading. The largest crypto asset, BTC, has a rather small number of days potentially affected by wash trading (8 or 0.51%). On the other hand, crypto assets with lower market cap and liquidity have a high number of days potentially infested with wash trading. This is consistent with the market microstructure theory, as the lower the market cap of a crypto asset, the easier its trading gets manipulated.

## 4.4 Comparisons of $\beta_{r_t}^{\ell}$ and $\beta_{\sigma_t}^{\ell}$ between Stocks and Cryptos

Comparing the descriptive statistics of liquidity jump $\beta_{r_t}^{\ell}$ for stocks (Panel A of Table 1) and crypto assets (Panel A of Table 2), we observe that for mean and median, as well as the number of days with high liquidity jump ($\beta_{r_t}^{\ell} \geq 1$), the values are numerically higher for crypto assets than for stocks by a wide margin (highest for stock: 38.27% for AMZN, lowest for crypto: 41.34% for XRP). This observation indicates that in general, crypto assets have higher liquidity fluctuation than stocks, which is consistent with the market-microstructure theory that low-liquidity assets (e.g., crypto assets) have higher liquidity fluctuation than high-liquidity assets (e.g., US stocks). A closer look shows that the five largest market-cap stocks (AAPL, AMZN, GOOG, MSFT and NVDA) also have the highest $\beta_{r_t}^{\ell}$ in both mean and median (mean: 1.20-1.39, median: 0.82-0.87), and the highest number of days with $\beta_{r_t}^{\ell}$ (30.32% to 38.27%). On the other hand, the three largest crypto assets in both market cap and liquidity (BTC, ETH and XRP) have the lowest $\beta_{r_t}^{\ell}$ in both mean and median (mean: 1.54, 1.63, 1.57, median: 0.92, 1.01, 0.85, respectively), and the lowest number of days with $\beta_{r_t}^{\ell} \geq 1$ (44.90%, 50.73%, 41.34%,).

---

[9] For the median of $\beta_{r_t}^{\ell}$ and the number of days with high $\beta_{r_t}^{\ell}$, both ETC and LTC have lower values than ETH does. However, this does not affect our subsequent discussions and results.

From comparisons between the descriptive statistics of liquidity diffusion $\beta_{\sigma_t}^{\ell}$ for stocks (Panel B of Table 1) and that for cryptos (Panel B of Table 2), we observe that for mean and median, the values are numerically higher for crypto assets than for stocks, indicating that crypto assets have higher liquidity volatility than stocks, which is consistent with the market microstructure noise theory. For the number of days with a high liquidity jump ($\beta_{\sigma_t}^{\ell} \geq 1$), the values are numerically higher for crypto assets than for stocks by a large margin, with a very noticeable exception of BTC, of which the value is 0.95% and is actually lower than that of NVDA at 1.09%. Again, the largest market-cap stocks (AAPL, AMZN, GOOG, MSFT and NVDA) have the highest $\beta_{\sigma_t}^{\ell}$ in both mean and median (mean: 0.76-0.79, median: 0.76-0.80) and the highest number of days with high $\beta_{\sigma_t}^{\ell}$ (0.23% to 1.09%) among stocks, and the three largest crypto assets (BTC, ETH and XRP) have the lowest number of days with high $\beta_{\sigma_t}^{\ell}$ (0.95%, 4.88%, 7.86%, respectively) among crypto assets.

In Subsection 4.2 we conjecture that wash trading occurs when both liquidity measures are high ($\beta_{r_t}^{\ell} \geq 1$ and $\beta_{\sigma_t}^{\ell} \geq 1$). For the stocks, there are between 0 and 15 or 0.57% days that exhibit both high $\beta^\ell_{r_t}$ and high $\beta^\ell_{\sigma_t}$, and its value is highest for the top five stocks (AAPL: 0.49%, AMZN: 0.41%, GOOG: 0.15%, MSFT: 0.38%, and NVDA: 0.57%). As the US stock market is heavily regulated and trading rules are rigorously monitored and enforced, it is essentially free of wash trading. As such, these top market-cap stocks establish the "top line" of stocks: those stocks with similar or lower number of days with $\beta^\ell_{r_t} \geq 1$ and $\beta^\ell_{\sigma_t} \geq 1$ are not affected by wash trading. For the crypto assets, the number of days with both high $\beta^\ell_{r_t}$ and high $\beta^\ell_{\sigma_t}$ has a rather wide span of 8 or 0.51% (BTC) to 545 or 34.56% (BCH). The largest crypto asset, BTC, has a rather small number of days affected by wash trading (0.51%), thus its theoretical exposure to wash trading is not more severe than that for large market-cap stocks, such as AAPL, AMZN and NVDA (0.53%, 0.57% and 1.09%, respectively). As such, BTC establishes the "bottom line" for crypto assets: those crypto assets with similar or higher number of days with $\beta^\ell_{r_t} \geq 1$ and $\beta^\ell_{\sigma_t} \geq 1$ are likely be infected by wash trading. By this criterion, the second and third largest crypto assets, ETH and XRP, have a relatively small number of days affected by wash trading (53 or 3.36%, 66 or 4.19%, respectively), while the other crypto assets are much more severely infected.

We observe a very narrow yet distinct overlap between "top line" stocks and "bottom line" crypto assets (i.e., comparing BTC with AAPL, AMZN, and NVDA). This finding reveals that, although the overall crypto market is fragmented, unregulated with shallow depth, the largest crypto assets (particularly BTC) traded in established (though still unregulated) exchanges such as Binance exhibit liquidity provision similar to that of the largest stocks, and therefore are moderately affected by wash trading. Comparing Figure 1 and Figure 2, we see that the distribution $\beta^\ell_{r_t}$-$\beta^\ell_{\sigma_t}$ for crypto assets is markedly different from that of stocks. Except for BTC and to a lesser extent ETH, all other examined crypto assets exhibit extreme liquidity diffusion ($\beta^{\ell}_{\sigma_t} > 3$), which indicates that crypto assets experience higher liquidity volatility than stocks, suggesting that the former are more prone to wash trading. We employ an ANOVA procedure to test the statistical significance of these estimates in Section 5.

### 4.5 Sensitivity Test with Cryptos $\beta^{\ell}_{r_t}$ and $\beta^{\ell}_{\sigma_t}$ with Treatment

To test the robustness of the conjecture that wash trading only occurs when both liquidity measures are high ($\beta^{\ell}_{r_t} \geq 1$ and $\beta^{\ell}_{\sigma_t} \geq 1$), we need to test the sensitivity of these liquidity measures to liquidity changes: if we intentionally remove a certain number of minute-level trades that are likely wash trades (essentially very high-volume trades in crypto markets), how will the values of $\beta^{\ell}_{r_t}$ and $\beta^{\ell}_{\sigma_t}$ change and by how much. In particular, different responses between $\beta^{\ell}_{r_t}$ and $\beta^{\ell}_{\sigma_t}$ to the removal of likely wash trades would help disclose the behavioral patterns of manipulative traders.

We repeat the procedure of deriving the daily $\beta^{\ell}_{r_t}$ and $\beta^{\ell}_{\sigma_t}$ for crypto assets, only this time with a "treatment" on wash trading, which is essentially a simulated "regulative measure" designed to remove extreme liquidity variability. Specifically, when we aggregate the tick-level data to construct minute-level amount ($A_\tau$), we divide $A_\tau$ into four quartiles (Q1-Q4) of equal quantity and reduce the quantity of Q3 (50 percentile) by a factor of 50%, and the quantity of Q4 (75 percentile) by a factor of 75%. By aggregating the minute-level data we derive the daily amount $A_t$, which is on average about 40% of the untreated "raw" daily amount across the selected crypto assets (or about 60% of the raw daily amount is regarded as being from wash trades). Since Cong et al. (2023) estimate that wash trades count for 46.47% of the total amount in Binance, our treatment is more stringent, making our results robust. We then calculate $\beta^{\ell}_{r_t}$ and $\beta^{\ell}_{\sigma_t}$ and report the descriptive statistics for all 10 crypto assets in Table 3.

Panels A and B of Table 3 summarize the descriptive statistics of the $\beta_{r_t}^\ell$ and $\beta_{\sigma_t}^\ell$ with treatment on wash trading, respectively. All the main measures are lower than their corresponding values for the same crypto assets without the treatment (Panels A and B of Table 2), indicating that treatment on wash trading indeed reduces liquidity fluctuation and liquidity volatility. However, the levels of reduction are quite different between $\beta_{r_t}^\ell$ and $\beta_{\sigma_t}^\ell$. In particular, the number of days with high $\beta_{r_t}^\ell$ ($\geq 1$) is between 502 or 31.83% (BTC) to 751 or 47.62% (UNI) with the treatment, while without the treatment, the number is between 652 or 41.34% (XRP) and 868 or 55.04% (SOL). The number of days with high $\beta_{\sigma_t}^\ell$ ($\geq 1$) is between 0 (BTC and ETH) to 154 or 9.77% (BCH) with the treatment, and the same measure is between 15 or 0.95% (BTC) and 968 or 61.38% (BCH) without the treatment. These results clearly point to that $\beta_{\sigma_t}^\ell$ is much more sensitive to the removal of likely wash trades than $\beta_{r_t}^\ell$. Furthermore, Panel C of Table 3 shows that the number of days with both high $\beta_{r_t}^\ell$ ($\geq 1$) and high $\beta_{\sigma_t}^\ell$ ($\geq 1$) with the treatment ranges from 0 (BTC, ETH and XRP) to 95 or 6.02% (BCH), which is significantly reduced from the same measure without the treatment at 8 or 0.51% (BTC) to 545 or 34.56% (BCH). Based on our conjuncture, with treatment, the potential exposure to wash trading for the top three crypto assets (BTC, ETH and XRP) is no more severe than that for the top stocks (AAPL, AMZN, GOOG, MSFT and NVDA).

Comparing the distribution of $\beta_{r_t}^\ell$-$\beta_{\sigma_t}^\ell$ in Figure 3 against that in Figures 1 and 2, we observe that the treatment essentially reduces the liquidity diffusion $\beta_{\sigma_t}^\ell$ of crypto assets to a level that is comparable to that of stocks. The most obvious observation is that a plot in Figure 3 (with the treatment) has a much narrower interval along the x-axis ($\beta_{\sigma_t}^\ell$) than its corresponding plot in Figure 2 (without the treatment) with the visible exceptions of BTC and ETH, while both plots have the same interval along the y-axis ($\beta_{r_t}^\ell$) with a maximum value of 10. These visualizations illustrate that treatment on wash trading results in a marginally lower number of days with a high liquidity jump and a significantly lower number of days with high liquidity diffusion. We employ an ANOVA procedure to test the statistical significance of the estimates in Section 5.

In addition to testing the robustness of our conjuncture that wash trading occurs when $\beta_{r_t}^{\ell} \geq 1$ and $\beta_{\sigma_t}^{\ell} \geq 1$, the sensitivity test also helps reveal how manipulative traders conduct wash trading. That $\beta_{\sigma_t}^{\ell}$ is much more sensitive than $\beta_{r_t}^{\ell}$ in responding to the treatment indicates that wash trading exhibits high liquidity diffusion ($\beta_{\sigma_t}^{\ell} \geq 1$) that reflects the manipulative traders' behavioral pattern of placing frequent and short-term momentum trades with drastically different price points on either bid or ask side, and wash trading also exhibits high liquidity jump ($\beta_{r_t}^{\ell} \geq 1$) that signifies the manipulative traders' timing preference to profit after having created high liquidity diffusion. Therefore, high liquidity diffusion "precedes" high liquidity jump, that manipulative traders "start" with stirring up high liquidity diffusion, and "end" at profiting from high liquidity jump. On the other hand, when liquidity diffusion is low ($\beta_{\sigma_t}^{\ell} < 1$), high liquidity jump ($\beta_{r_t}^{\ell} \geq 1$) actually indicates legitimate high-liquidity trades under either regulated (e.g., stocks) or unregulated (e.g., crypto assets) institutional settings. These findings are consistent with the behavioral dynamics demonstrated by a simple hypothetical trading environment consisting of both manipulative traders creating price distortions, and passive investors placing legitimate high-demand trades (see Appendix A for details).

## 5. ANOVA Tests for Robustness

### 5.1 ANOVA and Pairwise Tests

In the previous section, we establish that wash trading exhibits both high liquidity diffusion ($\beta_{\sigma_t}^{\ell} \geq 1$) and high liquidity jump ($\beta_{r_t}^{\ell} \geq 1$). We have three observations: 1) crypto assets without treatment on wash trading have higher liquidity jump and diffusion than stocks, and therefore the former are more likely exposed to wash trading, 2) a crypto asset with the treatment exhibits lower liquidity jump and diffusion than itself without the treatment, and therefore is exposed to lower level of wash trading, and 3) the largest market-cap crypto assets with the treatment exhibit much reduced levels of liquidity measures that are comparable to the largest market-cap stocks. Observations 2 and 3 indicate that our conjuncture is robust.

In order to examine the statistical significance of the above observations and validate our findings, we further conduct a series of ANOVA tests. First, we conduct a number of 2-asset/3-element 1-way ANOVA tests. Each ANOVA test is conducted on one liquidity measure (either $\beta_{r_t}^{\ell}$ or $\beta_{\sigma_t}^{\ell}$), using two assets with three elements: one stock, one crypto asset without treatment, and one crypto asset (the same crypto asset) with the treatment. For example, an ANOVA test of $\beta_{\sigma_t}^{\ell}$ on AAPL_BTC_wo_BTC_w produces the ANOVA properties for liquidity diffusion on stock AAPL and crypto asset BTC without and with the treatment, in which BTC_wo represents BTC without ('wo') the treatment, and BTC_w with ('w') the treatment. As there are 15 stocks and 10 crypto pairs (each crypto asset with and without the treatment), the combination yields 150 ANOVA tests for either $\beta_{r_t}^{\ell}$ or $\beta_{\sigma_t}^{\ell}$, or a total of 300 possible ANOVA tests.

If the result of an ANOVA test is significant ($p\text{-}value \leq 0.05$), i.e., at least one of the three means of the stock and crypto pair (e.g., AAPL_BTC_wo_BTC_w) is not statistically equal, we further conduct a set of pairwise tests on the same assets (e.g., AAPL_BTC_wo, AAPL_BTC_w, and BTC_wo_BTC_w, thus pairwise test consists of three 2-element "paired" tests). We repeat the same pairwise test three times, each with one of three different alternative hypotheses ("two-sided," "greater" or "less") to capture all statistically significant non-equalities. Thus, there are up to 450 (150 × 3) pairwise tests for each liquidity measure, or 900 possible pairwise tests.[10]

**5.2 Results and Discussions**

We find that all the 300 ANOVA tests produce significant results. As such, each 2-asset/3-element combination yields three pairwise tests with different alternatives ("two-sided", "greater" or "less"). Also, we find that each and every crypto asset without the treatment has significantly greater value than any stock for both liquidity measures, therefore we statistically validate our first observation that crypto assets without the treatment have higher liquidity jump and diffusion than stocks. Combining this finding with the observations that crypto assets without the treatment have a higher number of days with high values (≥ 1) for both liquidity measures than stocks (Panel C of Table 1 vs. Panel C of Table 2), we conclude, with statistical confidence, that crypto assets (without the treatment) are exposed to wash trading.

Furthermore, we find that every crypto asset without the treatment has significantly greater values than itself with the treatment for both liquidity measures, which statistically validates our second observation. Further combining this finding with that crypto assets without the treatment have higher number of days with wash trading than the same assets with the treatment (Panel C of Table 2 vs. Panel C of Table 3), we conclude, with statistical confidence, that the treatment is effective in removing wash trades, and therefore our conjuncture is robust and thus valid.[11]

---

[10] Note that for a pairwise test, 'greater than' is statistically equivalent to 'less than' if the result is significant (non-equality holds) for the paired assets. The reason we conduct pairwise tests with both 'greater than' and 'less than' alternatives is that Python actually compares the pairs in an alphabetical order, and therefore some pairs are only tested in 'greater than' alternative, while others only in 'less than' alternative. Therefore, 'greater than' and 'less than' tests are complementary and together they cover all possible pairs that are statistically unequal.

[11] In this paper, we do not report the results of ANOVA tests as they are all statistically significant. We do not report the results of the pairwise tests of a stock and a crypto without the treatment, or of a crypto without the treatment and the same asset with the treatment in the main manuscript to keep it concise. All these results are significant and are reported in supplement files and available if requested.

In order to establish the statistical confidence for the third observation, in Table 4, we report the results of all pairwise tests on one liquidity measure (either $\beta_{\sigma_t}^{\ell}$ or $\beta_{r_t}^{\ell}$) that is statistically no less for the stock than for the crypto asset with the treatment.[12] For liquidity jump $\beta_{r_t}^{\ell}$, we find that the values of two market-cap stocks (AMZN, GOOG) are statistically greater than that of BTC with the treatment, and the value of GOOG is statistically greater than that of XRP with the treatment (Panel A of Table 4). For liquidity diffusion $\beta_{\sigma_t}^{\ell}$, we find that the values of all top five market cap stocks (AAPL, AMZN, GOOG, MSFT, NVDA) are statistically greater than the top four market cap crypto assets (BTC, ETH, XRP), as well as SOL (except for AAPL) and LTC (Panel B of Table 4). These statistical results validate our early assertion that the treatment has a marginal effect in reducing the liquidity jump, but a major impact on reducing the liquidity diffusion. Also, we statistically validate our third observation that large-cap crypto assets with the treatment exhibit reduced levels of liquidity measures that are comparable to the large-cap stocks.

In Figure 4, we illustrate/visualize the liquidity measures of the top five stocks (AAPL, AMZN, GOOG, MSFT, NVDA) relative to the top three crypto assets (BTC, ETH, XRP) in a series of 2D point plots. These point plots clearly show that both liquidity measures of crypto assets without treatment are higher than that of some large market-cap stocks, while the liquidity measures of crypto assets with the treatment are comparable to those of some large market-cap stocks.

## 6. Conclusions

In this paper, we aim to quantify the size and probability of crypto wash trading to help market participants make informed trading decisions. We hypothesize that manipulated trades create liquidity fluctuations, which manifest as short-term price jumps that manipulative traders exploit for profit. To measure liquidity fluctuation, we propose that asset liquidity consists of two distinct yet complementary components: liquidity jump and liquidity diffusion (denoted as $\beta^{\ell}_{\sigma_t}$ and $\beta^{\ell}_{r_t}$, respectively). Liquidity jump is defined as the ratio of regular return to liquidity-adjusted return that captures the magnitude of daily liquidity fluctuations, and liquidity diffusion is the ratio of regular volatility to liquidity-adjusted volatility that reflects intraday liquidity volatility.

---

[12] In Table 4, we do not report the results of the pairwise tests of a stock and a crypto with treatment that are significant with the 'two-sided' alternative, nor do we report the results that are insignificant with the "greater than" or "less than" alternatives, for the reason of conciseness. These results are reported in supplement files and available upon request.

Using trading data from the U.S. stock market, which is highly regulated and therefore virtually immune to wash trading, as a benchmark, we conjecture that the combination of high liquidity diffusion and high liquidity jump ($\beta^{\ell}_{\sigma_t} \geq 1$, $\beta^{\ell}_{r_t} \geq 1$) is a reliable indicator of wash trading. Applying this framework to selected crypto assets, we find that, in general, crypto assets exhibit much higher liquidity diffusion and liquidity jump than stocks, suggesting exposure to wash trading. However, the distribution of liquidity measures suggests that wash trading is less prevalent among high-market-cap crypto assets traded on established though unregulated exchanges. In particular, the largest market-cap crypto assets, such as Bitcoin (BTC), exhibit relatively low level of wash trading. Conversely, lower-market-cap crypto assets are more susceptible to wash trading, which is consistent with theories of market microstructure and behavioral finance. From the perspective of investors, allocating capital to high-market-cap crypto assets (especially BTC) traded on established exchanges may help diversify portfolio risks.

To further validate our model, we test the robustness of these liquidity measures by examining their sensitivity to liquidity changes. We introduce a simulated regulatory treatment to filter out likely manipulative trades that contribute to extreme liquidity variability. Our results show that this treatment significantly reduces liquidity diffusion, while at the same time only marginally decreases liquidity jump. Notably, after applying the treatment, the largest market-cap crypto assets (BTC, ETH, and XRP) exhibit liquidity diffusion levels comparable to top market-cap stocks (e.g., AAPL, AMZN, GOOG, MSFT, and NVDA). The sensitivity tests suggest that high liquidity diffusion precedes high liquidity jump, that manipulative traders initially create excessive liquidity diffusion ($\beta_{\sigma_t}^{\ell} \geq 1$) in order to induce high liquidity jumps ($\beta_{r_t}^{\ell} \geq 1$), from which they profit. In contrast, large-volume trades that exhibit high liquidity jump but low liquidity diffusion (($\beta_{r_t}^{\ell} \geq 1$ and $\beta_{\sigma_t}^{\ell} < 1$) likely reflect legitimate high demand rather than manipulation. We present a theoretical model that further supports this interpretation by demonstrating that tactical manipulative trading simultaneously increases both the level and variance of liquidity pressure.

In summary, we provide answers to the questions raised in Section 1: 1) How should we adequately measure the size and probability of crypto wash trading on a daily basis? 2) What specific information should market participants analyze to make informed decisions in crypto trading? and 3) What targeted actions should regulators take to combat wash trading. The liquidity jump and liquidity diffusion form a liquidity Beta pair that maps well with the size and probability of crypto wash trading and provides a more comprehensive view of liquidity risk on daily basis. These liquidity measures provide a valuable tool for researchers, regulators, market makers, and investors to better understand the dynamics of liquidity variability in crypto trading. While we use crypto assets to exemplify the utilities of our model, the model can be extended to other asset classes characterized by extreme liquidity variability.

**Appendix A – Manipulative Trader vs. Passive Trader**

We consider a hypothetical trading environment over a trading day indexed by *t*, divided into *N* intervals (e.g., minutes). We consider two types of traders: 1) manipulative traders (M): trade to temporarily distort prices and profit from reversals, 2) passive investors (P): invest for exogenous reasons without strategized timing.

Suppose a manipulative trader places $N$ trades in day $t$, each inducing a distortion $\delta$ to the equilibrium return. For each of the $N$ trades, we write:

$$\mu_\tau = \mathbb{E}[\mu_\tau] + \delta$$

where:
$\mu_\tau$ is observed return at minute $\tau$, $\mathbb{E}[\mu_\tau]$ is the expected return at equilibrium
$\delta$ is the distortion purposely created by manipulative traders.

In order to create the distortion $\delta$, the manipulative trader incurs a cost $c(\delta)$ that is convex, e.g., $c(\delta) = \kappa \delta^2$. The manipulative trader chooses $\delta$ to maximize profit, given as:

$$\max_{\delta} \Pi_M = \delta - c(\delta) = \delta - \kappa \delta^2$$

$$d\Pi_M / d\delta = 1 - 2\kappa\delta = 0 \Rightarrow \delta^* = 1/2\kappa$$

Thus, an optimized manipulative trade induces a positive return distortion $\delta^*$, and if the practice is done repeatedly in day $t$, it induces an excessive daily return of $N\delta^*$ on day $t$, and increases the daily variance as well. As such, the series of manipulative trades leads to elevated size and variance of daily return, resulting in high values in both liquidity jump and liquidity diffusion ($\beta_{r_t}^\ell \geq 1$ and $\beta_{\sigma_t}^\ell \geq 1$).

On the other hand, a passive investor does not submit trades aiming at creating price distortion, therefore these trades do not induce volatility beyond normal microstructure noise. The return of a trade a passive investor places is expressed as:

$$\mu_\tau = \mathbb{E}[\mu_\tau] \Rightarrow \delta \to 0$$

where: $\mu_\tau$ is observed return at minute $\tau$, $\mathbb{E}[\mu_\tau]$ is the expected return at equilibrium

If the trades are indeed high volume (high liquidity), they are likely to cause significant price change on the daily base. Thus, while high-liquidity trades placed by a passive investor may cause increased daily return, they do not induce elevated volatility. As such, we may see high liquidity jump but low liquidity diffusion ($\beta^{\ell}_{r_t} \geq 1$ and $\beta^{\ell}_{\sigma_t} < 1$).

Therefore, the hypothetical trading environment supports our empirical framework tests with an economic intuition, that 1) if both liquidity measures are high ($\beta^{\ell}_{r_t} \geq 1$ and $\beta^{\ell}_{\sigma_t} \geq 1$), the trading day *t* is infested by wash trades, whereas 2) if liquidity jump is high but liquidity diffusion is low ($\beta^{\ell}_{r_t} \geq 1$ and $\beta^{\ell}_{\sigma_t} < 1$), the trading day *t* sees mostly legitimate high demand trades.

# References

Acharya, V.V. and Pedersen, L.H., 2005. Asset pricing with liquidity risk. *Journal of Financial Economics,* 77(2), pp.375-410.

Aït-Sahalia, Y., Mykland, P.A. and Zhang, L., 2005. How often to sample a continuous-time process in the presence of market microstructure noise. *Review of Financial Studies,* 18(2), pp.351-416.

Aloosh, A. and Li, J., 2024. Direct evidence of bitcoin wash trading. *Management Science*, published online: March 15, 2024.

Amihud, Y., 2002. Illiquidity and stock returns: cross-section and time-series effects. *Journal of Financial Markets,* 5(1), pp.31-56.

Amihud, Y. and Mendelson, H., 1986. Liquidity and stock returns. *Financial Analysts Journal,* 42(3), pp.43-48.

Amihud, Y., Mendelson, H. and Pedersen, L.H., 2005. Liquidity and asset prices. *Foundations and Trends® in Finance,* 1(4), pp.269-364.

Amiram, D., Cserna, B., Kalay, A. and Levy, A., 2019. The Information Environment, Volatility Structure, and Liquidity. *Columbia Business School Research Paper,* 15-62.

Anderson, T. G., Bollerslev, T. and Meddahi, N., 2011. Realized volatility forecasting and market microstructure noise. *Journal of Econometrics,* 160, pp.220-234.

Bekaert, G., Harvey, C.R. and Lundblad, C., 2007. Liquidity and expected returns: Lessons from emerging markets. *Review of Financial Studies,* 20(6), pp.1783-1831.

Bollerslev, T. and Todorov, V., 2023. The jump leverage risk premium. *Journal of Financial Economics,* 150, 103772.

Brennan, M.J. and Subrahmanyam, A., 1996. Market microstructure and asset pricing: On the compensation for illiquidity in stock returns. *Journal of Financial Economics,* 41(3), pp.441-464.

Chan, L.K. and Lakonishok, J., 1995. The behavior of stock prices around institutional trades. *The Journal of Finance,* 50(4), pp.1147-1174.

Chordia, T. and Subrahmanyam, A., 2004. Order imbalance and individual stock returns: Theory and evidence. *Journal of Financial Economics,* 72(3), pp.485-518.

Chordia, T., Subrahmanyam, A. and Anshuman, V.R., 2001. Trading activity and expected stock returns. *Journal of Financial Economics,* 59(1), pp.3-32.

Chung, K.H. and Zhang, H., 2014. A simple approximation of intraday spreads using daily data. *Journal of Financial Markets,* 17, pp.94-120.

Cong, L. W., Li, X., Tang, K. and Yang, Y., 2023. Crypto wash trading. *Management Science,* 69, pp.6427-6454.

Datar, V.T., Naik, N.Y. and Radcliffe, R., 1998. Liquidity and stock returns: An alternative test. *Journal of Financial Markets,* 1(2), pp.203-219.

Easley, D. and O'Hara, M., 1987. Price, trade size, and information in securities markets. *Journal of Financial Economics,* 19(1), pp.69-90.

Eleswarapu, V.R., 1997. Cost of transacting and expected returns in the Nasdaq market. *Journal of Finance,* 52(5), pp.2113-2127.

Fong, K.Y., Holden, C.W. and Trzcinka, C.A., 2017. What are the best liquidity proxies for global research? *Review of Finance,* 21(4), pp.1355-1401.

Gabaix, X., Gopikrishnan, P., Plerou, V. and Stanley, H.E., 2006. Institutional investors and stock market volatility. *The Quarterly Journal of Economics,* 121(2), pp.461-504.

Glosten, L.R. and Milgrom, P.R., 1985. Bid, ask and transaction prices in a specialist market with heterogeneously informed traders. *Journal of Financial Economics,* 14(1), pp.71-100.

Hou, K. and Moskowitz, T.J., 2005. Market frictions, price delay, and the cross-section of expected returns. *Review of Financial Studies,* 18(3), pp.981-1020.

Ho, T. and Stoll, H.R., 1981. Optimal dealer pricing under transactions and return uncertainty. *Journal of Financial Economics,* 9(1), pp.47-73.

Huberman, G. and Halka, D., 2001. Systematic liquidity. *Journal of Financial Research,* 24(2), pp.161-178.

Kyle, A.S., 1985. Continuous auctions and insider trading. *Econometrica: Journal of the Econometric Society,* pp.1315-1335.

Le Pennec, G., Fiedler, I. and Ante, L., 2021. Wash trading at cryptocurrency exchanges. *Finance Research Letters,* 43, p.101982.

Manahov, V., 2021. Cryptocurrency liquidity during extreme price movements: is there a problem with virtual money? *Quantitative Finance,* 21, pp.341-360.

Menkveld, A.J., 2013. High frequency trading and the new market makers. *Journal of Financial Markets,* 16(4), pp.712-740.

Pástor, Ľ. and Stambaugh, R.F., 2003. Liquidity risk and expected stock returns. *Journal of Political Economy,* 111(3), pp.642-685.

Pereira, J.P. and Zhang, H.H., 2010. Stock returns and the volatility of liquidity. *Journal of Financial and Quantitative Analysis,* 45(4), pp.1077-1110.

Petkova, R., Akbas, F. and Armstrong, W.J., 2011. The volatility of liquidity and expected stock returns. *Available at SSRN 1786991*.

Roll, R., 1984. A simple implicit measure of the effective bid-ask spread in an efficient market. *Journal of Finance,* 39(4), pp.1127-1139.

Shen, D., Urquhart, A., Wang, P., 2022. Bitcoin intraday time series momentum. *The Financial Review,* 57, pp.319-344.

Sifat, I., Tariq, S.A. and van Donselaar, D., 2024. Suspicious trading in nonfungible tokens (NFTs). *Information & Management,* 61(1), pp.103898.

Tripathi, A. and Dixit, A., 2020. Liquidity of financial markets: a review. *Studies in Economics and Finance,* 37(2), pp.201-227.

## Table 1 - Descriptive Statistics of Stocks

Panels A and B report the descriptive statistics of liquidity jump ($\beta_{r_t}^{\ell}$) and liquidity diffusion ($\beta_{\sigma_t}^{\ell}$), respectively, for each stock over the entire sample period. Panel C reports the number of days with both $\beta_{r_t}^{\ell}$ and $\beta_{\sigma_t}^{\ell}$ being no less than 1. The maximum value of both $\beta_{r_t}^{\ell}$ and $\beta_{\sigma_t}^{\ell}$ is capped at 10.

| Panel A | liquidity jump ($\beta_{r_t}^{\ell}$) | | | | | | | | | | | | | | |
|---|---|---|---|---|---|---|---|---|---|---|---|---|---|---|---|
| ticker | AAPL | AMZN | ATI | CMA | CRS | EME | GOOG | IBKR | LII | MLI | MSFT | NVDA | TPL | VFC | WSO |
| count | 2652 | 2652 | 2652 | 2652 | 2652 | 2652 | 2652 | 2652 | 2652 | 2652 | 2652 | 2652 | 2652 | 2652 | 2652 |
| mean | 1.20 | 1.32 | 1.09 | 1.09 | 0.99 | 0.83 | 1.39 | 1.09 | 0.82 | 0.90 | 1.26 | 1.25 | 0.44 | 1.06 | 0.71 |
| std | 1.55 | 1.69 | 1.74 | 1.77 | 1.74 | 1.54 | 1.88 | 1.79 | 1.54 | 1.72 | 1.65 | 1.64 | 1.06 | 1.61 | 1.36 |
| min | 0.00 | 0.00 | 0.00 | 0.00 | 0.00 | 0.00 | 0.00 | 0.00 | 0.00 | 0.00 | 0.00 | 0.00 | 0.00 | 0.00 | 0.00 |
| median | 0.83 | 0.87 | 0.57 | 0.58 | 0.47 | 0.39 | 0.85 | 0.55 | 0.38 | 0.39 | 0.83 | 0.82 | 0.20 | 0.61 | 0.33 |
| max | 10.00 | 10.00 | 10.00 | 10.00 | 10.00 | 10.00 | 10.00 | 10.00 | 10.00 | 10.00 | 10.00 | 10.00 | 10.00 | 10.00 | 10.00 |
| number of days $\beta_{r_t}^{\ell} = 10$ | 48 | 54 | 53 | 59 | 55 | 32 | 65 | 59 | 36 | 57 | 47 | 52 | 20 | 38 | 27 |
| as % of total number of days | 1.81% | 2.04% | 2.00% | 2.22% | 2.07% | 1.21% | 2.45% | 2.22% | 1.36% | 2.15% | 1.77% | 1.96% | 0.75% | 1.43% | 1.02% |
| number of days $\beta_{r_t}^{\ell} \geq 1$ | 804 | 1015 | 652 | 648 | 560 | 433 | 997 | 662 | 480 | 481 | 894 | 888 | 187 | 666 | 381 |
| as % of total number of days | 30.32% | 38.27% | 24.59% | 24.43% | 21.12% | 16.33% | 37.59% | 24.96% | 18.10% | 18.14% | 33.71% | 33.48% | 7.05% | 25.11% | 14.37% |
| number of days $\beta_{r_t}^{\ell} \leq 0.10$ | 46 | 69 | 221 | 202 | 282 | 303 | 95 | 221 | 357 | 342 | 77 | 64 | 612 | 171 | 385 |
| as % of total number of days | 1.73% | 2.60% | 8.33% | 7.62% | 10.63% | 11.43% | 3.58% | 8.33% | 13.46% | 12.90% | 2.90% | 2.41% | 23.08% | 6.45% | 14.52% |

| Panel B | liquidity diffusion ($\beta_{\sigma_t}^{\ell}$) | | | | | | | | | | | | | | |
|---|---|---|---|---|---|---|---|---|---|---|---|---|---|---|---|
| ticker | AAPL | AMZN | ATI | CMA | CRS | EME | GOOG | IBKR | LII | MLI | MSFT | NVDA | TPL | VFC | WSO |
| count | 2652 | 2652 | 2652 | 2652 | 2652 | 2652 | 2652 | 2652 | 2652 | 2652 | 2652 | 2652 | 2652 | 2652 | 2652 |
| mean | 0.76 | 0.79 | 0.63 | 0.61 | 0.50 | 0.46 | 0.77 | 0.60 | 0.45 | 0.44 | 0.76 | 0.76 | 0.22 | 0.63 | 0.41 |
| std | 0.08 | 0.08 | 0.15 | 0.13 | 0.12 | 0.11 | 0.07 | 0.14 | 0.12 | 0.12 | 0.07 | 0.10 | 0.08 | 0.12 | 0.11 |
| min | 0.39 | 0.30 | 0.07 | 0.07 | 0.10 | 0.11 | 0.31 | 0.11 | 0.08 | 0.08 | 0.14 | 0.14 | 0.07 | 0.08 | 0.08 |
| median | 0.77 | 0.80 | 0.65 | 0.63 | 0.51 | 0.47 | 0.77 | 0.61 | 0.45 | 0.44 | 0.76 | 0.77 | 0.21 | 0.65 | 0.41 |
| max | 1.79 | 1.21 | 1.07 | 1.11 | 0.91 | 0.86 | 1.08 | 1.01 | 1.04 | 0.83 | 1.59 | 1.92 | 0.98 | 1.01 | 0.74 |
| number of days $\beta_{\sigma_t}^{\ell} = 10$ | 0 | 0 | 0 | 0 | 0 | 0 | 0 | 0 | 0 | 0 | 0 | 0 | 0 | 0 | 0 |
| as % of total number of days | 0.00% | 0.00% | 0.00% | 0.00% | 0.00% | 0.00% | 0.00% | 0.00% | 0.00% | 0.00% | 0.00% | 0.00% | 0.00% | 0.00% | 0.00% |
| number of days $\beta_{\sigma_t}^{\ell} \geq 1$ | 14 | 15 | 1 | 2 | 0 | 0 | 6 | 1 | 1 | 0 | 10 | 29 | 0 | 1 | 0 |
| as % of total number of days | 0.53% | 0.57% | 0.04% | 0.08% | 0.00% | 0.00% | 0.23% | 0.04% | 0.04% | 0.00% | 0.38% | 1.09% | 0.00% | 0.04% | 0.00% |
| number of days $\beta_{\sigma_t}^{\ell} \leq 0.10$ | 0 | 0 | 5 | 10 | 2 | 0 | 0 | 0 | 1 | 3 | 0 | 0 | 25 | 5 | 3 |
| as % of total number of days | 0.00% | 0.00% | 0.19% | 0.38% | 0.08% | 0.00% | 0.00% | 0.00% | 0.04% | 0.11% | 0.00% | 0.00% | 0.94% | 0.19% | 0.11% |

| Panel C | $\beta_{r_t}^{\ell} \geq 1$ & $\beta_{\sigma_t}^{\ell} \geq 1$ | | | | | | | | | | | | | | |
|---|---|---|---|---|---|---|---|---|---|---|---|---|---|---|---|
| ticker | AAPL | AMZN | ATI | CMA | CRS | EME | GOOG | IBKR | LII | MLI | MSFT | NVDA | TPL | VFC | WSO |
| count | 2652 | 2652 | 2652 | 2652 | 2652 | 2652 | 2652 | 2652 | 2652 | 2652 | 2652 | 2652 | 2652 | 2652 | 2652 |
| number of days $\beta_{r_t}^{\ell}, \beta_{\sigma_t}^{\ell} \geq 1$ | 13 | 11 | 1 | 1 | 0 | 0 | 4 | 0 | 1 | 0 | 10 | 15 | 0 | 1 | 0 |
| as % of total number of days | 0.49% | 0.41% | 0.04% | 0.04% | 0.00% | 0.00% | 0.15% | 0.00% | 0.04% | 0.00% | 0.38% | 0.57% | 0.00% | 0.04% | 0.00% |

# Table 2 - Descriptive Statistics of Crypto Assets – without Treatment on Wash Trading

Panels A and B report the descriptive statistics of liquidity jump ($\beta^{\ell}_{r_t}$) and liquidity diffusion ($\beta^{\ell}_{\sigma_t}$), respectively, for each crypto asset without treatment on wash trading. Panel C reports the number of days with both $\beta^{\ell}_{r_t}$ and $\beta^{\ell}_{\sigma_t}$ being no less than 1. The maximum value of both $\beta^{\ell}_{r_t}$ and $\beta^{\ell}_{\sigma_t}$ is capped at 10.

| Panel A | | | | | liquidity jump ($\beta^{\ell}_{r_t}$) | | | | | |
|---|---|---|---|---|---|---|---|---|---|---|
| ticker | ADA_wo | AVAX_wo | BCH_wo | BTC_wo | ETC_wo | ETH_wo | LTC_wo | SOL_wo | UNI_wo | XRP_wo |
| count | 1577 | 1577 | 1577 | 1577 | 1577 | 1577 | 1577 | 1577 | 1577 | 1577 |
| mean | 1.88 | 1.98 | 1.99 | 1.54 | 1.88 | 1.63 | 1.82 | 1.93 | 1.99 | 1.57 |
| std | 2.39 | 2.51 | 2.53 | 2.03 | 2.41 | 2.02 | 2.45 | 2.37 | 2.49 | 2.12 |
| min | 0.00 | 0.00 | 0.00 | 0.00 | 0.00 | 0.01 | 0.00 | 0.00 | 0.00 | 0.00 |
| median | 1.00 | 1.05 | 1.04 | 0.92 | 0.99 | 1.01 | 0.90 | 1.09 | 1.06 | 0.85 |
| max | 10 | 10 | 10 | 10 | 10 | 10 | 10 | 10 | 10 | 10 |
| number of days $\beta^{\ell}_{r_t} = 10$ | 77 | 85 | 87 | 47 | 70 | 47 | 67 | 72 | 78 | 47 |
| as % of total number of days | 4.88% | 5.39% | 5.52% | 2.98% | 4.44% | 2.98% | 4.25% | 4.57% | 4.95% | 2.98% |
| number of days $\beta^{\ell}_{r_t} >= 1$ | 789 | 813 | 817 | 708 | 783 | 800 | 724 | 868 | 825 | 652 |
| as % of total number of days | 50.03% | 51.55% | 51.81% | 44.90% | 49.65% | 50.73% | 45.91% | 55.04% | 52.31% | 41.34% |
| number of days $\beta^{\ell}_{r_t} <= 0.10$ | 62 | 67 | 76 | 55 | 62 | 52 | 76 | 64 | 72 | 68 |
| as % of total number of days | 3.93% | 4.25% | 4.82% | 3.49% | 3.93% | 3.30% | 4.82% | 4.06% | 4.57% | 4.31% |

| Panel B | | | | | liquidity diffusion ($\beta^{\ell}_{\sigma_t}$) | | | | | |
|---|---|---|---|---|---|---|---|---|---|---|
| ticker | ADA_wo | AVAX_wo | BCH_wo | BTC_wo | ETC_wo | ETH_wo | LTC_wo | SOL_wo | UNI_wo | XRP_wo |
| count | 1577 | 1577 | 1577 | 1577 | 1577 | 1577 | 1577 | 1577 | 1577 | 1577 |
| mean | 0.97 | 1.10 | 1.15 | 0.82 | 1.12 | 0.86 | 0.92 | 1.01 | 1.15 | 0.87 |
| std | 0.27 | 0.52 | 0.32 | 0.07 | 0.26 | 0.07 | 0.13 | 0.46 | 0.36 | 0.12 |
| min | 0.55 | 0.32 | 0.58 | 0.43 | 0.55 | 0.52 | 0.65 | 0.65 | 0.67 | 0.62 |
| median | 0.91 | 0.94 | 1.06 | 0.82 | 1.05 | 0.85 | 0.90 | 0.88 | 1.05 | 0.85 |
| max | 6.17 | 6.01 | 5.13 | 1.15 | 2.92 | 1.26 | 3.49 | 8.42 | 6.01 | 3.86 |
| number of days $\beta^{\ell}_{\sigma_t} = 10$ | 0 | 0 | 0 | 0 | 0 | 0 | 0 | 0 | 0 | 0 |
| as % of total number of days | 0.00% | 0.00% | 0.00% | 0.00% | 0.00% | 0.00% | 0.00% | 0.00% | 0.00% | 0.00% |
| number of days $\beta^{\ell}_{\sigma_t} >= 1$ | 413 | 607 | 968 | 15 | 940 | 77 | 252 | 358 | 924 | 124 |
| as % of total number of days | 26.19% | 38.49% | 61.38% | 0.95% | 59.61% | 4.88% | 15.98% | 22.70% | 58.59% | 7.86% |
| number of days $\beta^{\ell}_{\sigma_t} <= 0.10$ | 0 | 0 | 0 | 0 | 0 | 0 | 0 | 0 | 0 | 0 |
| as % of total number of days | 0.00% | 0.00% | 0.00% | 0.00% | 0.00% | 0.00% | 0.00% | 0.00% | 0.00% | 0.00% |

| Panel C | | | | | $\beta^{\ell}_{r_t} >= 1$ & $\beta^{\ell}_{\sigma_t} >= 1$ | | | | | |
|---|---|---|---|---|---|---|---|---|---|---|
| ticker | ADA_wo | AVAX_wo | BCH_wo | BTC_wo | ETC_wo | ETH_wo | LTC_wo | SOL_wo | UNI_wo | XRP_wo |
| count | 1577 | 1577 | 1577 | 1577 | 1577 | 1577 | 1577 | 1577 | 1577 | 1577 |
| number of days $\beta^{\ell}_{r_t}, \beta^{\ell}_{\sigma_t} >= 1$ | 256 | 344 | 545 | 8 | 513 | 53 | 166 | 219 | 530 | 66 |
| as % of total number of days | 16.23% | 21.81% | 34.56% | 0.51% | 32.53% | 3.36% | 10.53% | 13.89% | 33.61% | 4.19% |

## Table 3 - Descriptive Statistics of Crypto Assets – with Treatment on Wash Trading

Panels A and B report the descriptive statistics of liquidity jump ($\beta_{r_t}^{\ell}$) and liquidity diffusion ($\beta_{\sigma_t}^{\ell}$), respectively, for each crypto asset with treatment on wash trading. Panel C reports the number of days with both $\beta_{r_t}^{\ell}$ and $\beta_{\sigma_t}^{\ell}$ being no less than 1. The maximum value of both $\beta_{r_t}^{\ell}$ and $\beta_{\sigma_t}^{\ell}$ is capped at 10.

| Panel A | | | | | liquidity jump ($\beta_{r_t}^{\ell}$) | | | | | |
|---|---|---|---|---|---|---|---|---|---|---|
| ticker | ADA_w | AVAX_w | BCH_w | BTC_w | ETC_w | ETH_w | LTC_w | SOL_w | UNI_w | XRP_w |
| count | 1577 | 1577 | 1577 | 1577 | 1577 | 1577 | 1577 | 1577 | 1577 | 1577 |
| mean | 1.51 | 1.56 | 1.59 | 1.21 | 1.61 | 1.37 | 1.43 | 1.54 | 1.63 | 1.26 |
| std | 2.06 | 2.10 | 2.09 | 1.68 | 2.13 | 1.89 | 1.95 | 2.16 | 2.12 | 1.82 |
| min | 0.00 | 0.00 | 0.00 | 0.00 | 0.00 | 0.01 | 0.00 | 0.00 | 0.00 | 0.00 |
| median | 0.84 | 0.89 | 0.90 | 0.75 | 0.86 | 0.82 | 0.81 | 0.85 | 0.95 | 0.74 |
| max | 10 | 10 | 10 | 10 | 10 | 10 | 10 | 10 | 10 | 10 |
| number of days $\beta_{r_t}^{\ell} = 10$ | 48 | 50 | 47 | 31 | 51 | 42 | 43 | 64 | 57 | 41 |
| as % of total number of days | 3.04% | 3.17% | 2.98% | 1.97% | 3.23% | 2.66% | 2.73% | 4.06% | 3.61% | 2.60% |
| number of days $\beta_{r_t}^{\ell} >= 1$ | 640 | 679 | 691 | 502 | 694 | 575 | 611 | 631 | 751 | 510 |
| as % of total number of days | 40.58% | 43.06% | 43.82% | 31.83% | 44.01% | 36.46% | 38.74% | 40.01% | 47.62% | 32.34% |
| number of days $\beta_{r_t}^{\ell} <= 0.10$ | 58 | 66 | 70 | 62 | 59 | 53 | 56 | 53 | 62 | 57 |
| as % of total number of days | 3.68% | 4.19% | 4.44% | 3.93% | 3.74% | 3.36% | 3.55% | 3.36% | 3.93% | 3.61% |

| Panel B | | | | | liquidity diffusion ($\beta_{\sigma_t}^{\ell}$) | | | | | |
|---|---|---|---|---|---|---|---|---|---|---|
| ticker | ADA_w | AVAX_w | BCH_w | BTC_w | ETC_w | ETH_w | LTC_w | SOL_w | UNI_w | XRP_w |
| count | 1577 | 1577 | 1577 | 1577 | 1577 | 1577 | 1577 | 1577 | 1577 | 1577 |
| mean | 0.77 | 0.80 | 0.86 | 0.66 | 0.85 | 0.70 | 0.75 | 0.76 | 0.86 | 0.70 |
| std | 0.09 | 0.10 | 0.11 | 0.05 | 0.10 | 0.04 | 0.07 | 0.09 | 0.10 | 0.06 |
| min | 0.39 | 0.35 | 0.40 | 0.36 | 0.40 | 0.43 | 0.46 | 0.45 | 0.50 | 0.40 |
| median | 0.75 | 0.78 | 0.85 | 0.67 | 0.84 | 0.70 | 0.75 | 0.73 | 0.85 | 0.70 |
| max | 1.34 | 1.45 | 1.34 | 0.85 | 1.39 | 0.88 | 1.49 | 1.25 | 1.55 | 1.58 |
| number of days $\beta_{\sigma_t}^{\ell} = 10$ | 0 | 0 | 0 | 0 | 0 | 0 | 0 | 0 | 0 | 0 |
| as % of total number of days | 0.00% | 0.00% | 0.00% | 0.00% | 0.00% | 0.00% | 0.00% | 0.00% | 0.00% | 0.00% |
| number of days $\beta_{\sigma_t}^{\ell} >= 1$ | 34 | 44 | 154 | 0 | 104 | 0 | 6 | 29 | 100 | 1 |
| as % of total number of days | 2.16% | 2.79% | 9.77% | 0.00% | 6.59% | 0.00% | 0.38% | 1.84% | 6.34% | 0.06% |
| number of days $\beta_{\sigma_t}^{\ell} <= 0.10$ | 0 | 0 | 0 | 0 | 0 | 0 | 0 | 0 | 0 | 0 |
| as % of total number of days | 0.00% | 0.00% | 0.00% | 0.00% | 0.00% | 0.00% | 0.00% | 0.00% | 0.00% | 0.00% |

| Panel C | | | | | $\beta_{r_t}^{\ell} >= 1$ & $\beta_{\sigma_t}^{\ell} >= 1$ | | | | | |
|---|---|---|---|---|---|---|---|---|---|---|
| ticker | ADA_w | AVAX_w | BCH_w | BTC_w | ETC_w | ETH_w | LTC_w | SOL_w | UNI_w | XRP_w |
| count | 1577 | 1577 | 1577 | 1577 | 1577 | 1577 | 1577 | 1577 | 1577 | 1577 |
| number of days $\beta_{r_t}^{\ell}, \beta_{\sigma_t}^{\ell} >= 1$ | 27 | 30 | 95 | 0 | 62 | 0 | 3 | 20 | 67 | 0 |
| as % of total number of days | 1.71% | 1.90% | 6.02% | 0.00% | 3.93% | 0.00% | 0.19% | 1.27% | 4.25% | 0.00% |

# Table 4 – ANOVA Pairwise Tests for Stocks and Crypto Assets with Treatment

Panel A reports the results of associated pair tests for liquidity jump ($\beta_{r_t}^{\ell}$) with the value of a stock being statistically greater than that of a crypto asset with treatment on wash trading. Panel B reports the results of associated pair tests for liquidity diffusion ($\beta_{\sigma_t}^{\ell}$) with the value of a stock being statistically greater than that of a crypto asset with treatment on wash trading.

| Panel A | Pairwise Test - Liquidity Jump $\beta_{r_t}^{\ell}$ - Stock - Crypto with Treatment on Wash Trading | | | | | | | | |
|---|---|---|---|---|---|---|---|---|---|
| Contrast | | A | B | T | dof | alternative | p-unc | p-corr | p-adjust | sig |
| | AMZN | BTC_w | 2.12 | 3312.19 | greater | 0.02 | 0.05 | holm | ** |
| liquidity jump $\beta_{r_t}^{\ell}$ | BTC_w | GOOG | -3.22 | 3598.79 | less | 0.00 | 0.00 | holm | *** |
| | GOOG | XRP_w | 2.18 | 3391.83 | greater | 0.01 | 0.04 | holm | *** |

| Panel B | Pairwise Test - Liquidity Diffusion $\beta_{\sigma_t}^{\ell}$ - Stock - Crypto with Treatment on Wash Trading | | | | | | | | |
|---|---|---|---|---|---|---|---|---|---|
| Contrast | A | B | T | dof | alternative | p-unc | p-corr | p-adjust | sig |
| | AAPL | BTC_w | 47.86 | 4220.16 | greater | 0.00 | 0.00 | holm | *** |
| | AAPL | ETH_w | 32.81 | 4136.09 | greater | 0.00 | 0.00 | holm | *** |
| | AAPL | LTC_w | 2.76 | 3866.46 | greater | 0.00 | 0.01 | holm | *** |
| | AAPL | XRP_w | 26.09 | 4083.34 | greater | 0.00 | 0.00 | holm | *** |
| | ADA_w | AMZN | -8.30 | 3111.46 | less | 0.00 | 0.00 | holm | *** |
| | AMZN | BTC_w | 63.65 | 4213.07 | greater | 0.00 | 0.00 | holm | *** |
| | AMZN | ETH_w | 49.63 | 4153.55 | greater | 0.00 | 0.00 | holm | *** |
| | AMZN | LTC_w | 16.33 | 3826.88 | greater | 0.00 | 0.00 | holm | *** |
| | AMZN | SOL_w | 12.93 | 3190.50 | greater | 0.00 | 0.00 | holm | *** |
| | AMZN | XRP_w | 40.68 | 4055.69 | greater | 0.00 | 0.00 | holm | *** |
| liquidity diffusion $\beta_{\sigma_t}^{\ell}$ | BTC_w | GOOG | -54.64 | 4137.10 | less | 0.00 | 0.00 | holm | *** |
| | ETH_w | GOOG | -39.29 | 4215.72 | less | 0.00 | 0.00 | holm | *** |
| | GOOG | LTC_w | 6.15 | 3607.19 | greater | 0.00 | 0.00 | holm | *** |
| | GOOG | SOL_w | 4.05 | 2967.00 | greater | 0.00 | 0.00 | holm | *** |
| | GOOG | XRP_w | 31.01 | 3882.99 | greater | 0.00 | 0.00 | holm | *** |
| | BTC_w | MSFT | -49.14 | 4144.63 | less | 0.00 | 0.00 | holm | *** |
| | ETH_w | MSFT | -33.38 | 4213.00 | less | 0.00 | 0.00 | holm | *** |
| | MSFT | XRP_w | 25.97 | 3896.73 | greater | 0.00 | 0.00 | holm | *** |
| | BTC_w | NVDA | -43.51 | 4163.55 | less | 0.00 | 0.00 | holm | *** |
| | ETH_w | NVDA | -29.67 | 3902.17 | less | 0.00 | 0.00 | holm | *** |
| | LTC_w | NVDA | -3.20 | 4160.55 | less | 0.00 | 0.00 | holm | *** |
| | NVDA | SOL_w | 1.74 | 3690.83 | greater | 0.04 | 0.12 | holm | ** |
| | NVDA | XRP_w | 24.30 | 4226.32 | greater | 0.00 | 0.00 | holm | *** |

\*\*\* - significant at the 1% level, \*\* - significant at the 5% level, \* significant at the 10% level.

Note: For a crypto asset, suffix '_w' stands for with treatment on wash trading. For example, 'BTC_w' means the liquidity measure (either $\beta_{r_t}^{\ell}$ or $\beta_{\sigma_t}^{\ell}$) for BTC with the treatment.

# Figure 1 – Distribution of Stocks ($\beta^\ell_{r_t}$-$\beta^\ell_{\sigma_t}$)

This figure provides scatter plots of liquidity jump ($\beta^\ell_{r_t}$) and liquidity diffusion ($\beta^\ell_{\sigma_t}$) for 15 US stocks over the entire sample period. The max values of $\beta^\ell_{r_t}$ and $\beta^\ell_{\sigma_t}$ are capped at 10. The stocks are AAPL, AMZN, ATI, CMA, CRS, EME, GOOG, IBKR, LII, MLI, MSFT, NVDA, TPL, VFC, and WSO. The x-axis is liquidity diffusion ($\beta^\ell_{\sigma_t}$) and the y-axis is the liquidity jump ($\beta^\ell_{r_t}$).

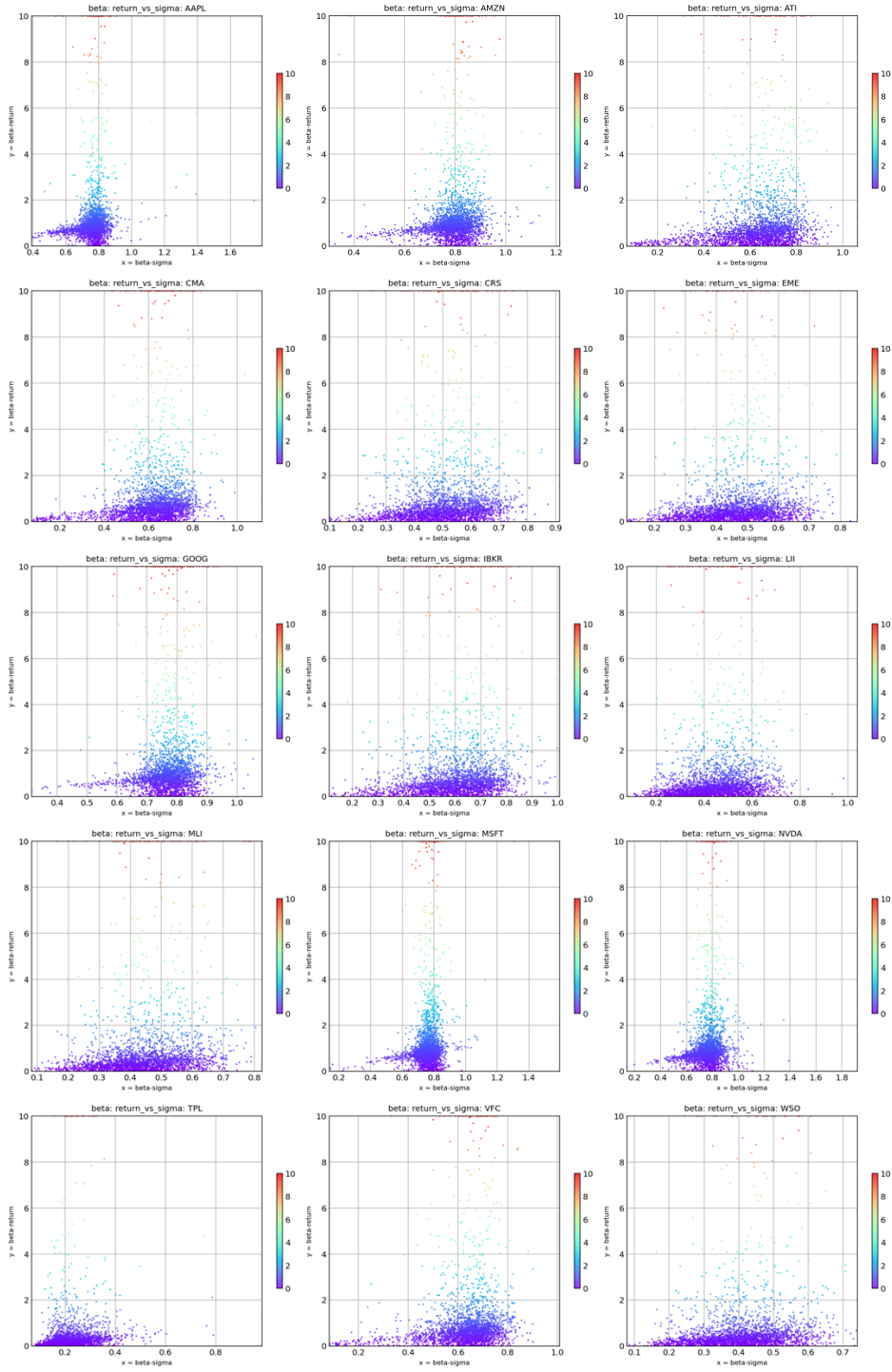

**Figure 2 – Distribution of Crypto Assets ($\beta^{\ell}_{r_t}$-$\beta^{\ell}_{\sigma_t}$) – without Treatment on Wash Trading**

This figure provides scatter plots of liquidity jump ($\beta^{\ell}_{r_t}$) and liquidity diffusion ($\beta^{\ell}_{\sigma_t}$) without treatment on wash trading for 10 crypto assets over the entire sample period. The max values of $\beta^{\ell}_{r_t}$ and $\beta^{\ell}_{\sigma_t}$ are capped at 10. The crypto assets are ADA, AVAX, BCH, BTC, ETC, ETH, LTC, SOL, UNI, and XRP. The x-axis is liquidity diffusion ($\beta^{\ell}_{\sigma_t}$) and the y-axis is the liquidity jump ($\beta^{\ell}_{r_t}$).

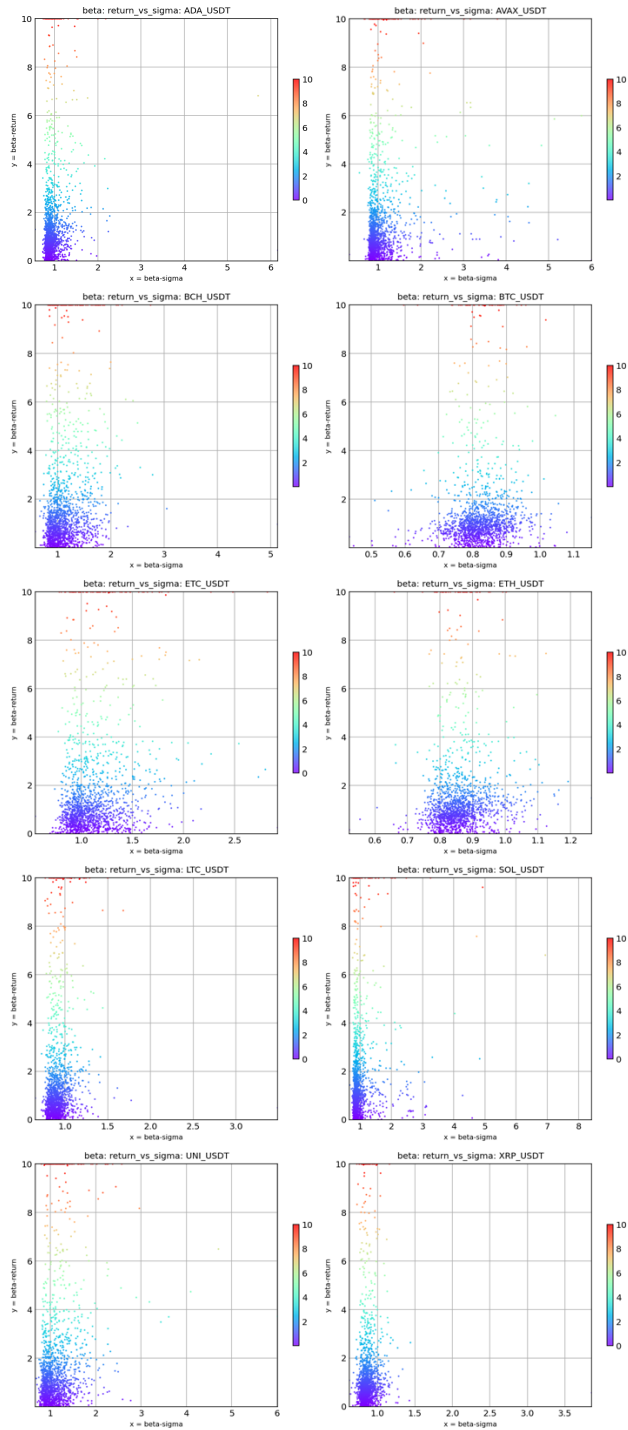

**Figure 3 – Distribution of Crypto Assets ($\beta^\ell_{r_t}$-$\beta^\ell_{\sigma_t}$) – with Treatment on Wash Trading**

This figure provides scatter plots of liquidity jump ($\beta^\ell_{r_t}$) vs. liquidity diffusion ($\beta^\ell_{\sigma_t}$) with treatment on wash trading for 10 crypto assets over the entire sample period. The max values of $\beta^\ell_{r_t}$ and $\beta^\ell_{\sigma_t}$ are capped at 10. The crypto asset ticker symbols are ADA, AVAX, BCH, BTC, ETC, ETH, LTC, SOL, UNI, and XRP. The x-axis is liquidity diffusion ($\beta^\ell_{\sigma_t}$) and the y-axis is the liquidity jump ($\beta^\ell_{r_t}$).

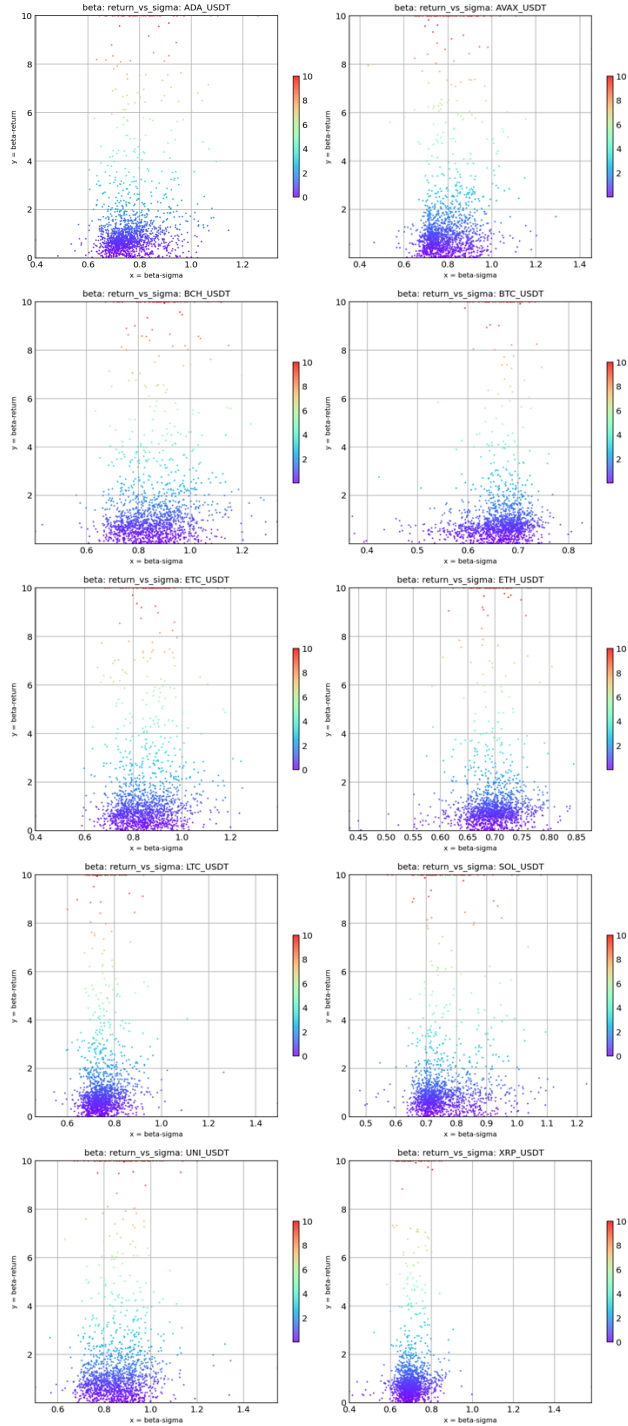

# Figure 4 – Point Plots of $\beta_{r_t}^{\ell}$-$\beta_{\sigma_t}^{\ell}$: ANOVA Tests

This figure provides point plots of $\beta_{r_t}^{\ell}$-$\beta_{\sigma_t}^{\ell}$ for one stock (AAPL, AMZN, GOOG, MSFT, NVDA) and one crypto asset (BTC, ETH, XRP) pair without and with treatment on wash trading. The x-axis is liquidity diffusion ($\beta_{\sigma_t}^{\ell}$) and the y-axis is the liquidity jump ($\beta_{r_t}^{\ell}$).

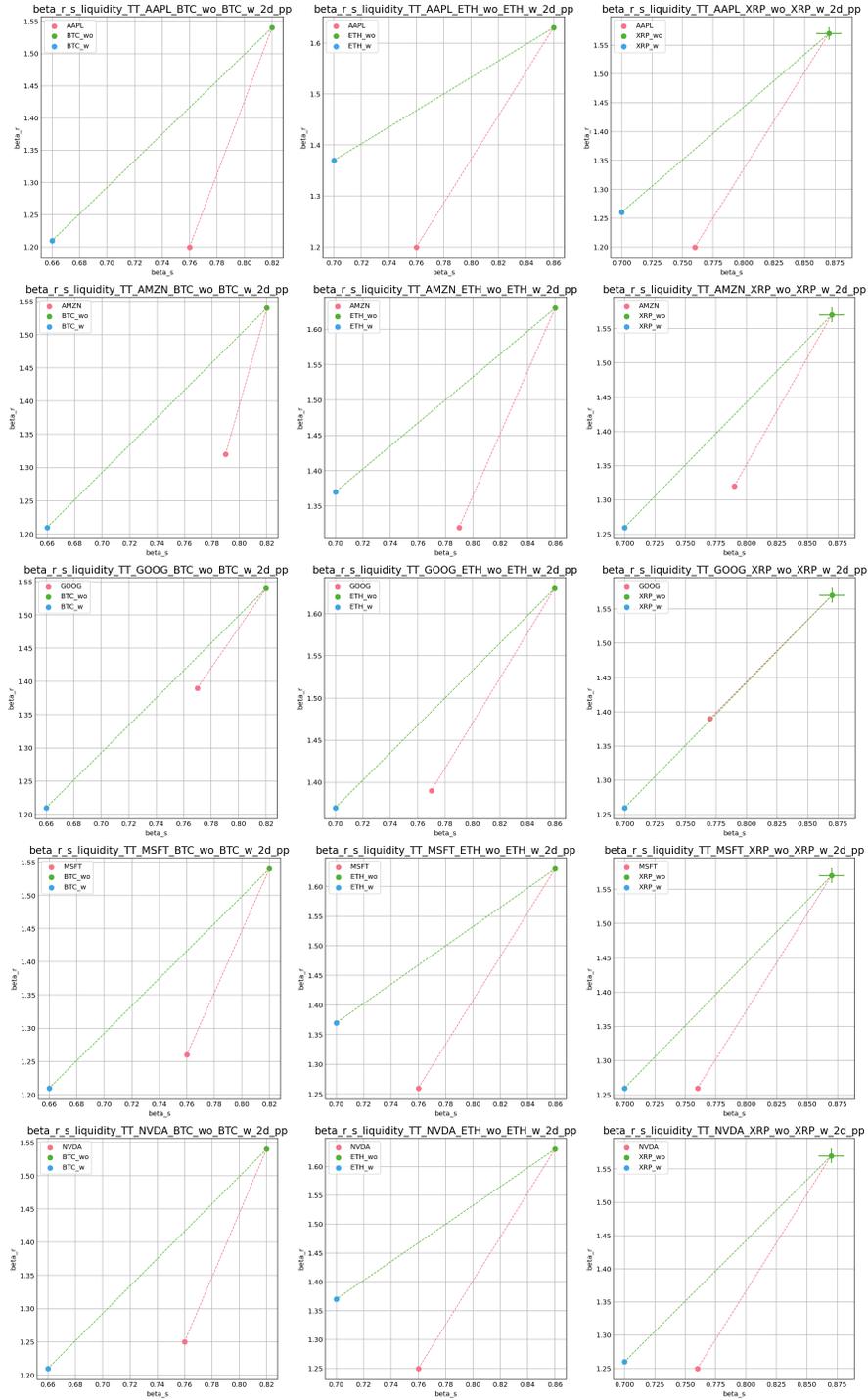